\documentclass[12pt, a4paper]{article}
\usepackage{amsmath,amssymb}
\usepackage{graphicx}
\usepackage{bm}
\usepackage{cite}
\usepackage{color}
\usepackage{url}
\usepackage{float} 
\usepackage{booktabs}
\usepackage[normalem]{ulem}
\definecolor{blue}{rgb}{0.27, 0.42, 0.81}
\usepackage[
colorlinks=true,
urlcolor=blue,
citecolor=blue,
linkcolor=red,
linktocpage=true,
]{hyperref}

\newcommand{\nn}{\nonumber}
\newcommand{\ord}{{\cal O}}

\newcommand{\A}{{\cal A}}
\def\mLb{{m_{\Lambda_b}}}
\def\mmLb{{m^2_{\Lambda_b}}}
\def\mL{{m_{\Lambda}}}
\def\mmL{{m^2_{\Lambda}}}

\def\re{{\rm Re}}  \def\im{{\rm Im}}

\def\ARpe#1{{A^R_{\perp_{#1}}}}  \def\ARpa#1{{A^R_{\|_{#1}}}}

\def\AsRpe#1{{A^{\ast R}_{\perp_{#1}}}}  \def\AsRpa#1{{A^{\ast R}_{\|_{#1}}}}

\def\cL{{\cal L}}

\def\g{\gamma}
\def\l{\lambda}

\def\LbLmm{{\Lambda_b\to\Lambda \mu^+\mu^-}}
\def\LbLll{{\Lambda_b\to\Lambda \ell^+\ell^-}}
\def\LbLnpll{{\Lambda_b\to\Lambda(\to p\pi^-)\ell^+\ell^-}}
\def\LbLnp{{\Lambda_b\to\Lambda(\to p\pi^-)\mu^+\mu^-}}

\allowdisplaybreaks
\begin{document}
\sloppy
\begin{center}
\vspace*{5cm}
{\LARGE\bf
\boldmath
$\Lambda_b\rightarrow \Lambda(\to p \pi^-) \ell^+\ell^-$ as probe of
CP-violating New Physics
}

\vspace*{1.2cm}
{\bf Diganta Das $^{a}$, Jaydeb Das $^{b}$, Girish Kumar $^{c}$ and Niladribihari Sahoo $^{d}$}
\\\vspace*{.5cm}
$^{a}$ Center for Computational Natural Sciences and Bioinformatics, International Institute of Information
Technology, Hyderabad 500 032, India\\
\vspace*{.2cm}
$^{b}$ Department of Physics and Astrophysics, University of Delhi, Delhi 110007, India\\
\vspace*{.2cm}
$^{c}$ Department of Physics, National Taiwan University, Taipei 10617,
Taiwan\\
\vspace*{.2cm}
$^{d}$ School of Physics and Astronomy, University of Birmingham, Birmingham B15 2TT, UK
\end{center}

\vspace*{5mm}
\begin{abstract}
\noindent
We investigate
the possible sizes of all the CP-violating asymmetries
offered by the angular distribution of rare decay
$\Lambda_b\rightarrow \Lambda(\to p \pi^-) \ell^+\ell^-$ in the
Standard Model and new physics scenarios motivated by the recent
$b\to s \ell^+\ell^-$ anomalies.
We work in a model-independent effective theory framework and
discuss the sensitivity of CP asymmetries to new ${O}_{9,10}$ operators and their
chirality flipped counterparts. We find that the size of
many of the CP asymmetries can be at the level of a few percent in new physics scenarios
consistent with current $b\to s\ell^+\ell^-$ data at a level of $1\sigma$.
We emphasize that measurements of these CP asymmetries can be used
to discriminate different new physics scenarios~in~$b\to s \ell^+\ell^-$.

\end{abstract}
\newpage
\section{Introduction} \label{sec: intro}

A major motivation for going beyond the Standard Model (SM)
is to find new sources of charge-parity violation (CPV)
required for the explanation of  baryonic asymmetry of the Universe (BAU).
The SM possesses two CPV sources: the Cabibbo-Kobayashi-Maskawa (CKM) phase
(related to weak interactions), and a strong CP
phase that is severely constrained by the upper limit of the neutron electric dipole moment measurement \cite{Abel:2020pzs}.
So far, all experimental observations of CPV  have been
in the quark sector and are consistent with the CKM mechanism.
However, it is well known that the SM fails to satisfy the
Sakharov's conditions \cite{Sakharov:1967dj} needed for
explaining the BAU. 
Therefore, investigations of experimentally accessible
CP-violating observables  offering excellent sensitivity to physics beyond
the SM are highly motivated.

At the luminosity frontier  of new physics (NP) searches,
the physical processes with underlying quark current
$b \to s \ell^+\ell^- $ transitions have been of 
particular interest in recent times.
The measurements of several observables related
to  processes $B \to K^{(\ast)} \mu^+\mu^-$
\cite{LHCb:2013ghj,LHCb:2014cxe,LHCb:2015svh,LHCb:2020lmf},
$B_s \to \phi \mu^+\mu^-$
\cite{LHCb:2015wdu,LHCb:2021xxq}, show deviations from
the SM expectation.
On the other hand,
the recently reported measurements of lepton flavor universality (LFU) ratios
$R_{K^{(\ast)}}={\cal B}(B\to K^{(\ast)}\mu^+\mu^-)/{\cal B}(B\to K^{(\ast)}e^+e^-)$
\cite{Hiller:2003js,Bobeth:2007dw} by the LHCb \cite{LHCb:2022qnv,LHCb:2022zom}, which are updates of previous measurements \cite{LHCb:2014vgu,LHCb:2017avl,LHCb:2019hip,LHCb:2021trn},
agree with the SM.
However, the latest
global likelihood analyses of $b \to s \ell^+\ell^-$ that include the latest 
$R_{K^{(\ast)}}$ measurements still show large preference for the NP hypothesis over the SM
\cite{SinghChundawat:2022zdf,Greljo:2022jac,Alguero:2023jeh}\footnote{Earlier fits can be found, for example, in Refs.~\cite{Descotes-Genon:2013wba,Altmannshofer:2013foa,Beaujean:2013soa,
Hurth:2013ssa,Hurth:2014vma,Altmannshofer:2014rta,Descotes-Genon:2015uva,
Capdevila:2017bsm,DAmico:2017mtc,Alguero:2019ptt,Kowalska:2019ley,Alok:2019ufo,
Hurth:2020ehu,Geng:2021nhg,Altmannshofer:2021qrr,Alguero:2021anc}.}
(also see Ref.~\cite{Ciuchini:2022wbq},
which discusses impact of long-distance contributions associated with charm loops). Although these results indicate the presence of a lepton flavor universal NP, it is also worth mentioning that there may still be sufficient room for LFU violation if the NP is associated with CP violation \cite{Fleischer:2023zeo}.
These anomalies, collectively known as neutral-current $B$-anomalies,
will be tested rigorously in the upcoming measurements with more data,
and, if confirmed, would be  indisputable evidence of NP
in $b \to s\ell^+\ell^-$ transitions.

The aforementioned anomalies, if confirmed, say nothing about the CP nature of the underlying NP as the concerned $b\to s\ell^+\ell^-$ observables are CP averaged.
However, $b \to s \ell^+ \ell^-$ decays also offer a multitude of observables 
which are highly sensitive to the CP nature of NP, aided by the fact that in the SM the $b\to s\ell^+\ell^-$ transitions are doubly Cabibbo suppressed \cite{Bobeth:2008ij}.
Combined measurements of
CP-violating and CP-conserving observables therefore provide a more powerful method
to understand the CP properties of NP.
In the literature, several works have investigated the CP asymmetries in
$B \to K^\ast \mu^+\mu^-$ angular distributions to probe NP
\cite{Kruger:1999xa,Kim:2007fx,Bobeth:2008ij,Altmannshofer:2008dz,Alok:2011gv,
Alok:2017jgr,Becirevic:2020ssj,SinghChundawat:2022zdf,Gangal:2022ole}, and, as illustrated in Ref.~\cite{Alok:2017jgr}, the
CP asymmetries  are capable of distinguishing between
NP models that address the $b \to s \ell^+\ell^-$ anomalies.

If the observed deviation in the $b\to s \ell^+ \ell^-$ is indeed due to
unambiguous short-distance NP then it would, in principle, also affect all
semileptonic processes with 
the same underlying current. 
One important example is the baryonic decay
$\Lambda_b \to  \Lambda (\to p \pi^-) \ell^+\ell^-$ for an unpolarized $\Lambda_b$,
which is the topic of this paper.
There are several benefits of studying this decay.
The angular distribution of $\Lambda_b \to  \Lambda (\to p \pi^-) \ell^+\ell^-$,
similar to its mesonic counterpart
$B \to K^\ast(\to K \pi) \ell^+ \ell^-$,
offers a large number of CP-conserving as well as CP-violating angular asymmetries
that provide complementary information about NP in
$b \to s \ell^+\ell^-$ \cite{Boer:2014kda}.
The secondary decay $K^\ast \to K \pi$ in mesonic mode $B \to K^\ast(\to K \pi) \ell^+\ell^-$
is a strong decay and therefore it conserves
parity. On the other hand, the decay $\Lambda\to p\pi^-$ in the baryonic mode is a
parity violating weak decay; this characteristic will play a key role in constructing several 
CP-violating asymmetries in $\Lambda_b \to  \Lambda (\to p \pi^-) \ell^+\ell^-$,
as discussed later in this paper.
Furthermore, as pointed out in Ref. \cite{Detmold:2016pkz},
$\Lambda_b \to \Lambda$ form factors, in comparison to $B \to K^\ast$,
are more suitable to be computed with higher precision using lattice QCD due to the 
stability of $\Lambda$ under strong interactions.

The first observation of
$\Lambda_b \to  \Lambda (\to p \pi^-) \ell^+\ell^-$ was reported by the
CDF \cite{CDF:2011buy}. The recent angular analysis of
unpolarized\footnote{A full angular analysis of $\Lambda_b \to  \Lambda (\to p \pi^-) \ell^+\ell^-$
with polarized $\Lambda_b$ has been performed
by LHCb in Ref.~\cite{LHCb:2018jna}. Predictions in the SM and
beyond for polarized $\Lambda_b$ decay $\Lambda_b \to  \Lambda (\to p \pi^-) \ell^+\ell^-$ can be found in Refs.~\cite{Blake:2017une,Das:2020qws}.} $\Lambda_b \to  \Lambda (\to p \pi^-) \ell^+\ell^-$
by the LHCb indicates a branching ratio that is smaller
than the SM expectation \cite{LHCb:2015tgy},
a pattern also observed in the $B\to K^{(\ast)}\mu^+\mu^-$, $B_s\to \phi\mu^+\mu^-$ modes.
There are extensive theoretical works on the model-independent study of
$\Lambda_b \to  \Lambda (\to p \pi^-) \ell^+\ell^-$
in the SM and beyond
\cite{Huang:1998ek,Chen:2001sj,Chen:2001ki,Aliev:2002nv,Aliev:2002hj,Aliev:2002tr,
Aliev:2004af,Aliev:2004yf,Aliev:2005np,Wang:2008sm,Aliev:2010uy,Mott:2011cx,
Gutsche:2013pp,Boer:2014kda,Kumar:2015tnz,Detmold:2016pkz,Meinel:2016grj,Faustov:2017wbh,
Blake:2017une,Roy:2017dum,Das:2018sms,Das:2018iap,Das:2019omf,
Bhattacharya:2019der,Blake:2019guk,Bordone:2021usz}.
These works mostly focus on CP-conserving angular observables.
To the best of our knowledge, angular observables which discriminate
the decay with its CP-conjugated mode
$\bar \Lambda_b \to \bar \Lambda (\to \bar p \pi^+) \ell^+\ell^-$ and their role in
the probe of NP in $b \to s \ell^+ \ell^-$ have not been discussed yet.
This paper investigates the prospects of CP asymmetries
associated with the baryonic mode and assesses their sensitivity to CP-violating NP.
We give description of all possible CP-violating asymmetries that are at disposal from
$\Lambda_b \to \Lambda (\to p \pi^-) \ell^+\ell^-$ angular distribution. Focusing on the muonic mode,
we identify the sensitivity of these asymmetries to various NP WCs and show that
the measurement of CP asymmetries in $\Lambda_b \to \Lambda (\to p \pi^-) \mu^+\mu^-$  can be used to distinguish
various NP solutions that are favored by current~global~$b\to s\mu^+\mu^-$~data.

The paper is organized as follows. We begin the next section with a description of
the model-independent effective framework relevant for the study of $b \to s\ell^+ \ell^-$ transitions.
In section~\ref{sec: LbLll} we define the full angular distribution of both
$\LbLnp$ and its CP-conjugated mode, and discuss the subtleties the CP properties of 
the secondary decay of $\Lambda\, (\bar \Lambda)$ particles bring in defining the
corresponding angular coefficients. In section \ref{sec: results}
we define all the  CP-violating asymmetries of the $\Lambda_b$
decay angular distribution and present our main numerical results.
Finally, we offer our conclusions~in~section~\ref{sec: conclusion}.

\section{$b \to s \ell^+\ell^-$ effective Lagrangian}\label{sec: eft}
In the SM, the  $b\to s \ell^+\ell^-$
 decays arise at the loop level only.
The Lagrangian relevant at the scale $\mu=m_b$ is given by \cite{Altmannshofer:2008dz},
\begin{align}\label{eq: Lag-b2sll}
	\cL_{\rm eff}
	= \frac{4G_F}{\sqrt{2}}\l_t\Big[
	C_1 O_1^c + C_2 O_2^c + \sum_{i=3}^{6} C_i O_i+ \sum_{i = 7}^{10}
	\left(C_i O_i+ C_i^\prime O_i^\prime\right)\nn\\
	 + (\lambda_u/\lambda_t)\{C_1 (O_1^c-O_1^u) + C_2 (O_2^c-O_2^u)\}\Big]\,,
\end{align}
where $G_F$ is the Fermi's constant,
and $\l_i=V_{ib}V_{is}^\ast$, where $V_{ij}$ denotes the CKM matrix element. The  WCs $C_i$ 
contain information about short-distance physics associated with
local operators $O_i$.
For the discussion of the $b\to s\ell^+\ell^-$ transition,
$O_{7,9,10}$ are the dominant operators which we list below.
Denoting the chiral projectors as
$P_{L,R} = (1\mp \g_5)/2$, one has,
\begin{align}\label{eq: O7-10}
O_7 &= \frac{e}{16\pi^2}m_b [\bar s \sigma_{\mu\nu} P_R b]F^{\mu\nu}, \\
O_{9} &= \frac{e^2}{16\pi^2}\left[\bar s\, \g_\mu P_L b\right]\left[\bar \ell\,\g^\mu \ell\right], 
\quad
O_{10} = \frac{e^2}{16\pi^2}\left[\bar s\, \g_\mu P_L b\right]\left[\bar \ell\,\g^\mu
\g_5 \ell\right],\label{eq: O9-10}
\end{align}
with $C_7(m_b) \simeq -0.3$, and $C_9(m_b)\approx - C_{10}(m_b)\simeq  4.2$
in the SM. Note that the primed operators are the chirality flipped counterparts
of the SM operators,
and in the SM their contributions $(C_i^\prime)$ are vanishing.
The operators $O_{1-6}$ are 4-quark operators related to decays
$b \to s \bar q q$, and $O_8$ is dipole operator
related to radiative decay $b \to s \gamma$. Their explicit form can be
found, for example, in
Refs.~\cite{Grinstein:1988me,Buchalla:1995vs,Chetyrkin:1996vx}.
These operators contribute to the $b \to s \ell^+\ell^-$ through quark loops.
It is customary to include their contribution in
the effective WCs $C_{7 (9)}^{\rm eff}$ of operators $O_{7 (9)}$.
These effective coefficients at
next-to-next-to-leading logarithmic are given as,
\begin{align}\label{eq: C7eff}
C_7^{\rm eff}   &= C_7 - \frac{1}{3}\left(C_3 + \frac{4}{3}C_4 + 20 C_5 + \frac{80}{3}C_6 \right) - \frac{\alpha_s}{4 \pi}\left((C_1-6 C_2)F_{1, c}^{7}(q^2) + C_8 F_8^{7}(q^2)\right),\\
C_9^{\rm eff} &= C_9 + \frac{4}{3} C_3 + \frac{64}{9}C_5 + \frac{64}{27}C_6
+ h(q^2, 0)\left(-\frac{1}{2}C_3 -\frac{2}{3}C_4 - 8C_5 - \frac{32}{3}C_6\right) \nn\\
& + h(q^2, m_b)\left(-\frac{7}{2}C_3 - \frac{2}{3}C_4 - 38 C_5 - \frac{32}{3} C_6\right) + h(q^2, m_c)\left(\frac{4}{3}C_1 + C_2 + 6 C_3 + 60 C_5\right) \nn\\
& - \frac{\alpha_s}{4 \pi}\left(C_1 F_{1,c}^{9}(q^2)+C_2F_{2, c}^{9}(q^2) +C_8 F_8^{9}(q^2)\right) + \frac{\l_u}{\l_t}\left(\frac{4}{3}C_1 + C_2\right)\left(h(q^2, m_c) - h(q^2, 0)\right).
\label{eq: C9eff}
\end{align}
The functions $h(a, b)$ and $F_8^{7, 9}(q^2)$ are given in Ref.~\cite{Beneke:2001at},
and the functions $F_{i, c}^{7, 9}(q^2)$ $(i =1, 2)$ are provided in
Refs.~\cite{Asatryan:2001zw,Greub:2008cy}. Note that the quark masses appearing in
Eqs.~\eqref{eq: C7eff} and \eqref{eq: C9eff} are in the pole scheme,
and the corresponding values ($m_b^{\rm pole} = 4.74174$ GeV,
$m_c^{\rm pole} = 1.5953$ GeV) are taken from Ref.~\cite{Detmold:2016pkz}. 
The numerical values of SM Wilson coefficients contributing to $b\to s \ell^+\ell^-$
are given in Table~\ref{Table:WC}.

 \begin{table}[h!]
 \begin{center}
	\setlength{\tabcolsep}{3pt}
	\begin{tabular}{ c | c | c | c | c | c | c | c | c | c  }
		\hline\hline
		$C_1$  & $C_2$ & $C_3$ & $C_4$ & $C_5$ &$C_6$ & $C_7$ & $C_8$ & $C_9$ & $C_{10}$  \\
		\hline
		$-0.2877$ & $1.0101$ & $-0.0060$ & $-0.0860$ & $0.0004$ & $0.0011$ & $-0.3361$ & $-0.1821$ & $4.2745$ & $-4.1602$\\
		\hline\hline
	\end{tabular}
	 \end{center}
	\caption{Values of $b\to s \ell^+\ell^-$ WCs in the SM at $\mu=4.2$ GeV taken from Ref.~\cite{Detmold:2016pkz}.}  
	\label{Table:WC}
\end{table}

There are $(\bar s b)(\bar \ell \ell )$ operators
with scalar and tensor structures which may arise in
NP; but  scalar operators are highly constrained
from $B_s\to\mu^+\mu^-$ data \cite{Alonso:2014csa, Beaujean:2015gba}.
Furthermore, since the current global fits to $b\to s \mu^+ \mu^-$ data
strongly prefer NP in WCs of left-handed (axial)vector operators in Eq.~\eqref{eq: O9-10},
we will also neglect NP tensor operators for the simplicity of the analysis.

\section{Angular Distribution} \label{sec: LbLll}
Assuming the $\Lambda_b$ to be unpolarized, the 4-fold differential angular distribution for $\LbLnpll$ is given by
\cite{Boer:2014kda},
\begin{align}\label{eq: ang-dist}
\frac{\mathrm{d}^{4} \Gamma}{\mathrm{d} q^{2} \mathrm{d} \cos \theta_{\ell}
\mathrm{d}\cos \theta_{\Lambda} \mathrm{d} \phi}=\frac{3}{8 \pi}[&\left(K_{1 s
s} \sin ^{2} \theta_{\ell}+K_{1 c c} \cos ^{2} \theta_{\ell}+K_{1 c} \cos
\theta_{\ell}\right)\nonumber \\
&+\left(K_{2 s s} \sin ^{2} \theta_{\ell}+K_{2 c c} \cos ^{2}
\theta_{\ell}+K_{2 c} \cos \theta_{\ell}\right) \cos \theta_{\Lambda}\nonumber \\
&+\left(K_{3 s c} \sin \theta_{\ell} \cos \theta_{\ell}+K_{3 s} \sin
\theta_{\ell}\right) \sin \theta_{\Lambda} \sin \phi\nonumber \\
&\left.+\left(K_{4 s c} \sin \theta_{\ell} \cos \theta_{\ell}+K_{4 s} \sin
\theta_{\ell}\right) \sin \theta_{\Lambda} \cos \phi\right].
\end{align}

The distribution is completely described by four variables: the invariant lepton mass squared ($q^2$)
and three Euler angles, $\theta_\ell$, $\theta_\Lambda$, and $\phi$. In the rest frame of $\Lambda_b$, the daughter baryon is assumed to travel along the $+z$ axis. The 
$\theta_\Lambda$ is the angle made by the proton with the $+z$ axis in 
the rest frame of the $\Lambda$,
$\theta_\ell$ is the angle made by the $\ell^-$ with respect to the $+z$ axis in
the rest frame of the lepton pair,
and $\phi$ defines angle in the rest frame of $\Lambda_b$ between planes
containing $p\pi^-$ and the lepton pair. 
Denoting mass of $\Lambda_b$, $\Lambda$, and charged lepton $\ell$ as
$m_{\Lambda_b}$, $m_\Lambda$, and $m_\ell$, respectively,
the physical region of the decay process is defined by the following values:
\begin{align}
	q^2 \in [4 m_\ell^2, (m_{\Lambda_b}-m_\Lambda)^2], \quad  \cos\theta_\Lambda \in [-1, 1],
	\quad  \cos\theta_\ell \in [-1, 1], \quad  \phi \in [0, 2\pi].
\end{align}

In Eq.~\eqref{eq: ang-dist}, the angular coefficients $K_i$  are functions of $q^2$.
These are conveniently described in terms of
$\Lambda_b\to\Lambda$ transversity amplitudes, $A^{L, R}_i(q^2)$ as follows,
\begin{align}\label{eq: coeff-K-first}
K_{1ss} &= \frac{1}{4} \big( 2|\ARpa0|^2 + |\ARpa1|^2 + 2|\ARpe0|^2 + |\ARpe1|^2 + \{ R \leftrightarrow L  \} \big), \\
K_{1cc} &= \frac{1}{2}\big( |\ARpa1|^2 + |\ARpe1|^2 + \{R \leftrightarrow L \} \big) ,\\
K_{1c} &= - \re\big( A^R_{\perp_1}A^{\ast R}_{\|_1} - \{ R \leftrightarrow L \}  \big),\\
K_{2ss} &= \frac{\alpha_\Lambda}{2} \re\big( 2 A^R_{\perp_0}A^{\ast R}_{\|_0} + A^R_{\perp_1}A^{\ast R}_{\|_1} + \{ R \leftrightarrow L \} \big) ,\\
K_{2cc} &= \alpha_\Lambda \re \big( A^R_{\perp_1}A^{\ast R}_{\|_1} + A^L_{\perp_1}A^{\ast L}_{\|_1} \big) ,\\
K_{2c} &= -\frac{\alpha_\Lambda}{2} \re \big( |\ARpe1|^2 + |\ARpa1|^2 - \{ R \leftrightarrow L \}   \big),\\
K_{3sc} &= \frac{\alpha_\Lambda}{\sqrt{2}} \im\big( \ARpe1\AsRpe0 - \ARpa1\AsRpa0 + \{ R \leftrightarrow L \} \big),\\
K_{3s} &= \frac{\alpha_\Lambda}{\sqrt{2}} \im\big( A^R_{\perp_1}A^{\ast R}_{\|_0} - A^R_{\|_1}A^{\ast R}_{\perp_0} - \{ R \leftrightarrow L \} \big),\\
K_{4sc} &= \frac{\alpha_\Lambda}{\sqrt{2}} \re\big( A^R_{\perp_1}A^{\ast R}_{\|_0} - A^R_{\|_1}A^{\ast R}_{\perp_0} + \{ R \leftrightarrow L \} \big),\\
K_{4s} &= \frac{\alpha_\Lambda}{\sqrt{2}} \re \big( A^R_{\perp_1}A^{\ast R}_{\perp_0} - A^R_{\|_1}A^{\ast R}_{\|_0} - \{ R \leftrightarrow L \} \big),
\label{eq: coeff-K-end}
\end{align}
where the $q^2$ dependence of the transversity amplitudes is implied. Since we will be focusing on the muonic mode only, we have
ignored the lepton mass in writing the coefficients $K_i$ above.
The expressions for $K_i$, including lepton mass effects, can be found in
Ref.~\cite{Das:2018iap}.  The transversity amplitudes $A^{L, R}_i(q^2)$ in terms of the $\Lambda_b \to \Lambda$
form factors $f_i(q^2)$ (see section \ref{sec: results}) and $b \to s\ell^+\ell^-$ Wilson coefficients are given by \cite{Das:2018iap}
\begin{align}
A^{L,(R)}_{\perp_1} &= -\sqrt{2}N \bigg( f^V_\perp \sqrt{2s_-} C^{L,(R)}_{ +} + \frac{2m_b}{q^2} f^T_\perp (\mLb + \mL) \sqrt{2s_-} C_7^{\rm eff} \bigg)\, ,\\
A^{L,(R)}_{\|_1} &= \sqrt{2}N \bigg( f^A_\perp \sqrt{2s_+} C^{L,(R)}_{-} + \frac{2m_b}{q^2} f^{T5}_\perp (\mLb - \mL) \sqrt{2s_+} C_7^{\rm eff} \bigg)\, ,\\
A^{L,(R)}_{\perp_0} &= \sqrt{2}N \bigg( f^V_0 (\mLb + \mL) \sqrt{\frac{s_-}{q^2}} C^{L,(R)}_{+} + \frac{2m_b}{q^2} f^T_0\sqrt{q^2s_-} C_7^{\rm eff} \bigg)\, ,\\
A^{L,(R)}_{\|_0} &= -\sqrt{2}N \bigg( f^A_0 (\mLb - \mL) \sqrt{\frac{s_+}{q^2}} C^{L,(R)}_{-} + \frac{2m_b}{q^2} f^{T5}_0\sqrt{q^2s_+} C_7^{\rm eff} \bigg)\, ,
\end{align}
where  $s_\pm = (\mLb \pm \mL)^2 - q^2$, and $C_{\pm}^{L(R)}$ are 
combinations of Wilson coefficients,
\begin{align}
 C_{+}^{L(R)} = (C_9^{\rm eff}\mp C_{10})+(C_{9}^\prime \mp C_{10}^\prime)\, ,\quad 
 C_{-}^{L(R)} = (C_9^{\rm eff}\mp C_{10})-(C_{9}^\prime \mp C_{10}^\prime)\, ,
\end{align}
and  the normalization constant $N$, a function of $q^2$, is given by,
\begin{equation}
N(q^2) =  \left[ \frac{G_F^2 |V_{tb}V_{ts}^\ast|^2 \alpha_e^2}{3.2^{11} m^3_{\Lambda_b} \pi^5 } {{q^2{\lambda^{1/2}(\mmLb,\mmL,q^2)}}{}\beta_\ell}\right]^{1/2}, \quad \beta_\ell = \sqrt{1 - \frac{4m_\ell^2}{q^2}}\,,
\end{equation}
with $\lambda(a,b,c) = a^2 + b^2 + c^2 -2(ab + bc + ca)$. 

In the expressions of $K_i$, the parity violating decay parameter $\alpha_\Lambda$ arises through the secondary decay $\Lambda\to p\pi^-$. The corresponding
hadronic matrix element is given by \cite{Boer:2014kda},
\begin{align}\label{eq: L2ppi matrix}
	\langle p (k_1, s_p) \pi^-(k_2)| (\bar d  u)_{V-A} (\bar u s)_{V-A}|\Lambda(k, s_\Lambda)\rangle
	= [\bar u(k_1, s_p)(\xi\, \gamma_5 + \omega)u (k, s_\Lambda)]\,,
\end{align}
which depends on only two parameters,  $\xi$ and $\omega$. These can be
determined from experimental data on $\Lambda \to p \pi^-$.
The decay parameter $\alpha_\Lambda$ then is given by \cite{Boer:2014kda},
\begin{align}\label{eq: alpha}
\alpha_\Lambda = \frac{-2\, \text{Re}(\omega\, \xi )}{\sqrt{r_{-}/r_{+}}|\xi|^2 + \sqrt{r_{+}/r_{-}}|\omega|^2}\,.
\end{align}
where $r_\pm = (m_\Lambda\pm m_p)^2 - m_\pi^2$.

To write down the corresponding distribution for CP-conjugated mode
$\bar \Lambda_b \to \bar \Lambda (\to \bar p \pi^+) \ell^+\ell^-$,
we take the following definition for Euler angles: 
In the rest frame of the $\bar{\Lambda_b}$, the $\bar{\Lambda}$ is assumed to travel along the $+z$ axis, and $\theta_\Lambda$ is the angle between $\bar \Lambda$ and the antiproton in the 
$(\bar p\pi^+)$ rest frame. The lepton angle
$\theta_\ell$ is the angle between $\bar \Lambda$ and $\ell^-$\footnote{
Note that direction of $\ell^-$ is taken as reference for both decay
and CP-conjugated mode. This convention is similar to the one used for mesonic counterpart decay in
Ref.~\cite{Altmannshofer:2008dz}.} in the dilepton rest frame,
and $\phi$ is the angle between planes of $(\bar p\pi^+)$ and the lepton pair. With the above convention, the decay distribution for
$\bar \Lambda_b \to \bar \Lambda (\to \bar p \pi^+) \ell^+\ell^-$
is simply obtained from Eq.~\eqref{eq: ang-dist} after the transformation
($\theta_\Lambda \to \theta_\Lambda,$ $\theta_\ell = \theta_\ell -\pi$, and $\phi \to -\phi $).
This is equivalent to replacing the functions $K_i$'s in Eq.~\eqref{eq: ang-dist}  with $ \bar K_i$'s
in the following way:
\begin{align}\label{eq: rule1}
	K_{1cc, 1ss, 2cc, 2ss, 4sc, 3s} \to &  + \bar K_{1cc, 1ss, 2cc, 2ss, 4sc, 3s}\,,\\
	K_{1c, 2c, 4s, 3sc} \to& - \bar K_{1c, 2c, 4s, 3sc}\,,
\end{align}
where $\bar K_i$ equals $K_i$ except for the weak phases conjugated. Additionally, in all but  three coefficients, $K_{1ss}$, $K_{1cc}$, and $K_{1c}$, we replace the $\Lambda\to p \pi^-$ decay parameter $\alpha_\Lambda$ by $\bar\alpha_\Lambda$, which corresponds to the CP-conjugated decay $\bar\Lambda\to \bar p \pi^+$. This replacement follows from the fact that under a CP transformation the Dirac-field bilinears transform as
$\bar \psi_1 \psi_2 \xrightarrow{CP}\bar \psi_2\psi_1$, $\bar \psi_1 \gamma_5\psi_2 \xrightarrow{CP} - \bar \psi_2 \gamma_5\psi_1$.
Therefore, in the expression of $\bar\Lambda\to \bar p \pi^+$, the hadronic matrix element obtained through a
CP transformation of equation \eqref{eq: L2ppi matrix},  the $\xi$ term picks up a minus sign
but the $\omega$ term does not.
The definition of $\bar\alpha_\Lambda$, defined similarly to Eq.~\eqref{eq: alpha},
then implies $\bar \alpha_\Lambda = -\alpha_\Lambda$ under strict CP-symmetry.
Experimentally measured values  $\alpha_\Lambda = 0.7519 \pm 0.0036 \pm 0.0024$
and $\bar\alpha_\Lambda=-0.7559 \pm 0.0036\pm0.0030$
by   BESIII collaboration \cite{BESIII:2022qax}
agree well with theory.

\section{CP-Violating Asymmetries and Results} \label{sec: results}
Let $\Gamma(H\to f)$ be the decay rate of $H \to f$ (where $H$ indicates the initial state $\Lambda_b$ and $f$ indicates the final state) and $\bar\Gamma(\bar H\to \bar f)$ be the decay rate of the CP-conjugate mode $\bar H\to \bar f$. As one needs two interfering decay amplitudes to observe a direct CPV, we write the amplitudes as
\begin{align}
    A[H \to f] &= |A_1| e^{i\phi^w_1} e^{i\phi^s_1} + |A_2| e^{i\phi^w_2} e^{i\phi^s_2}\, ,\\
    A[\bar H \to \bar f] &= |A_1| e^{-i\phi^w_1} e^{i\phi^s_1} + |A_2| e^{-i\phi^w_2} e^{i\phi^s_2}\,,
\end{align}
where $\phi^w_{1,2}$ are the weak phases, $\phi^s_{1,2}$ are the strong phases (arising due to final state interactions), and $|A_{1,2}|$ are the moduli of the interfering matrix elements. The decay rate asymmetry that signals the presence of CPV is
\begin{equation}
    \Gamma(H \to f) - \bar\Gamma(\bar H \to \bar f) \propto |A_1||A_2| \sin(\phi_1^s-\phi_2^s)\sin(\phi_1^w-\phi_2^w)\,.
\end{equation}
As evident from this expression, a nonvanishing CP asymmetry requires that
both the relative strong and weak phases of the amplitudes
must be nonvanishing.
The SM has a finite strong phase emanating from the imaginary part of
$C_9^{\rm eff}$, which is generated by the
$\bar q q$ loops ($q=c, u$) in the current-current operators.
However, the weak phase, coming from the CKM elements in the
last term of Eq.~\eqref{eq: C9eff}, is doubly Cabibbo suppressed and small. 
Therefore, the CPV in $b\to s\ell^+\ell^-$ transitions in the SM is expected
to be very small.

As discussed in the Introduction, the measurements of several observables associated
with the $b \to s\ell^+\ell^-$ current are in tension with the SM,
and the global fits to data show large preference to NP hypothesis over the SM.
Except for the direct CP asymmetry $A_{\rm CP}^{K^{(\ast)}}$ \cite{LHCb:2014mit} and a few angular CP-asymmetries in $B\to K^\ast\mu^{+}\mu^{-}$ \cite{LHCb:2015svh} and $B_s\to \phi \mu^{+}\mu^{-}$ \cite{LHCb:2015wdu,LHCb:2021xxq}, most of the measured $b\to s\ell^+\ell^-$ observables, including the ones that show the tensions with the SM, are CP averaged and therefore are not sensitive to the complex phases of NP. 
Therefore whether the NP in question is real or complex is not clear at present. To answer this question,
one needs to study CP-violating observable in $b\to s\ell^+\ell^-$ transition. To this purpose, we construct several
CP-violating observables in $\LbLll$ decay and investigate their sensitivity to complex NP Wilson coefficients.
We follow the results of Ref.~\cite{SinghChundawat:2022zdf}
which performed a global fit analysis of the
$b\to s \mu^+\mu^-$ data to complex Wilson coefficients assuming lepton flavor universal couplings to electrons and muons. 
In particular, we  consider the following two LFU NP scenarios\footnote{Since
Ref.~\cite{SinghChundawat:2022zdf} provided only $1\sigma$ range of WC values,
in choosing NP benchmark for our analysis, we take median for the real part
but the maximum of the $1\sigma$ range for the imaginary part to have maximum CP violation
effects.}:
\begin{itemize}
	\item Case I: $C_9^\text{NP} = -1.07 - 0.86i$ (NP in left-handed vector operator only)
	\item Case II: $C_9^\text{NP} = - C_{10}^\text{NP} =  -0.58 - 1.21i$ ($SU(2)_L$ invariant NP)
\end{itemize}
associated with the left-handed operators only. In addition, we will also present
results for a  NP scenario involving purely right-handed current
\begin{itemize}
	\item Case III: $C_9^\prime = - C_{10}^\prime = 0.04 + 0.16i$ (Right-handed NP)
\end{itemize}
The Case I and Case II can significantly improve the theory description of the data.
The Case III, although cannot explain tensions in $b\to s$ data
(see Ref.~\cite{Carvunis:2021jga}), we include it in our analysis  to assess the
sensitivity of CP asymmetries to right-handed NP\footnote{
Combination $C_9=-C_9^\prime$ involving both left- and -right current NP is also favored \cite{SinghChundawat:2022zdf} by the current data.}.

In order to make numerical predictions of observables, one also needs $\Lambda_b\to\Lambda$ form factors for which we use lattice QCD results of
Ref.\cite{Detmold:2016pkz}.
The $\Lambda_b \to \Lambda$ hadronic matrix elements are parametrized in terms of ten form factors\footnote{Note that the form factors labels used in Ref. \cite{Detmold:2016pkz} are different from labels we use in this paper. These are related as $f^V_{t,0,\perp} = f_{0,+,\perp}$, $f^A_{t,0,\perp} = g_{0,+,\perp}$, $f^T_{0,\perp}=h_{+,\perp}$ and $f^{T5}_{0,\perp}=\tilde{h}_{+,\perp}$.}  $f^V_{t,0,\perp}$, $f^A_{t,0,\perp}$, $f^T_{0,\perp}$, $f^{T5}_{0,\perp}$, which are function of $q^2$.
The lattice calculations are fitted to two $z$-parametrizations: the ``nominal" fit and ``higher-order" fit.
Defining the parameter $z(q^2,t_+)$ as,
\begin{equation}
	z(q^2,t_+) = \frac{ \sqrt{t_+-q^2} - \sqrt{t_+-t_0}  }{ \sqrt{t_+-q^2} + \sqrt{t_+-t_0} }\, ,
\end{equation}
where $t_0 = (\mLb-\mL)^2$ and $t_+=(m_B+m_K)^2$, in the so called ``nominal" fit the form factor parametrization is given as,
\begin{equation}\label{eq:ff-nominal}
f(q^2) = \frac{1}{1-q^2/(m^{f}_{\rm pole})^2} \big[ a^{f}_0 + a_1^{f} z(q^2,t_+) \big]\, ,
\end{equation}
while in ``higher-order" fit, the parametrization is given as,
\begin{equation}\label{eq:ff-higher}
f(q^2) = \frac{1}{1-q^2/(m^{f}_{\rm pole})^2} \big[ a^{f}_0 + a_1^{f} z(q^2,t_+) + a_2^{f} (z(q^2,t_+))^2 \big]\, ,
\end{equation}
where values of the coefficients $a_i^f$ and the correlations among them are taken from
Ref.~\cite{Detmold:2016pkz}.

In our numerical analysis, apart from already discussed  form factors and the Wilson coefficients, we use  results of  CKMfitter Group~\cite{Charles:2004jd} for values of CKM elements, while particle masses and their lifetime values are taken from Particle Data Group~\cite{ParticleDataGroup:2022pth}.

With the numerical inputs at our disposal, we make bin-wise predictions of different observables that we describe in the next section. To be precise, the prediction of an observable  ${\cal O}(q^2)$ in a given bin $q^2\in [a, b]$ is
\begin{align}
	\langle {\cal O}\rangle = \frac{1}{|b-a|}{\int_a^b dq^2 \,{\cal O}}(q^2)\,.
\end{align} 
For an observable involving a ratio of two quantities, the binned prediction is obtained after integrating the numerator and denominator separately and then taking their ratio.
The two regions of $q^2$ where we make the predictions are $q^2\in [0.1, 6]$ and $q^2\in [15, 20]$. To avoid the charmonium resonances, we refrain from making any prediction in the $6 - 15~{\rm GeV}^2$ range.

Our numerical determinations are subject to uncertainties coming from different inputs including the form factors. Regarding form factor uncertainties, as described in Ref.~\cite{Detmold:2016pkz}, we use the ``nominal'' fit of
Eq.~\eqref{eq:ff-nominal} and for an estimation of systematic uncertainties
``higher-order'' fit of Eq.~\eqref{eq:ff-higher} is used. The total uncertainty is obtained after adding statistical and systematic uncertainties in quadrature. In the appendix \ref{app: predictions} we have collected bin-wise predictions of all observables that we describe next.

\subsection{CP asymmetry in decay rate}
The CP asymmetry in decay rate is defined as,
\begin{align}\label{eq: acp}
	{\cal A}_{CP} = \frac{d \Gamma/dq^2 - d \bar\Gamma/dq^2 }
						  {d \Gamma/dq^2 + d \bar\Gamma/dq^2}\,,
\end{align}
where $d \Gamma/dq^2$ and $d \bar\Gamma/dq^2$ are decay rates of the decay $\LbLnp$ and
its CP-conjugate decay respectively, which in terms of angular coefficients
are defined as,
\begin{align}\label{eq: dgamma_dq2}
	\frac{d \Gamma}{dq^2 } = 2 K_{1ss} + K_{1cc},\quad 
	\frac{d \bar \Gamma}{dq^2 } = 2 \bar K_{1ss} + \bar K_{1cc}\,.
\end{align}

\begin{figure}[h]
\center
\includegraphics[width=0.48\textwidth]{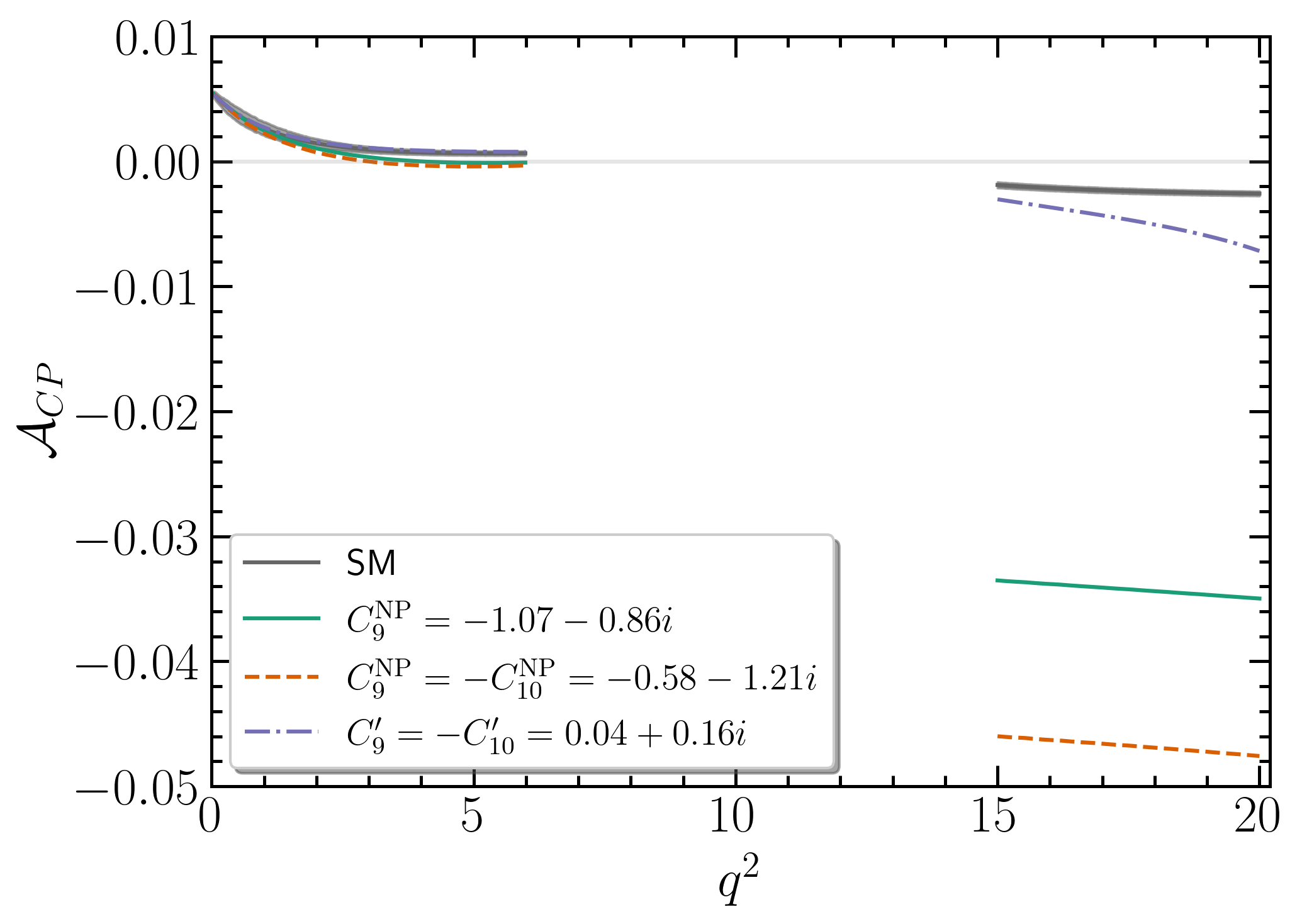}
\includegraphics[width=0.5\textwidth]{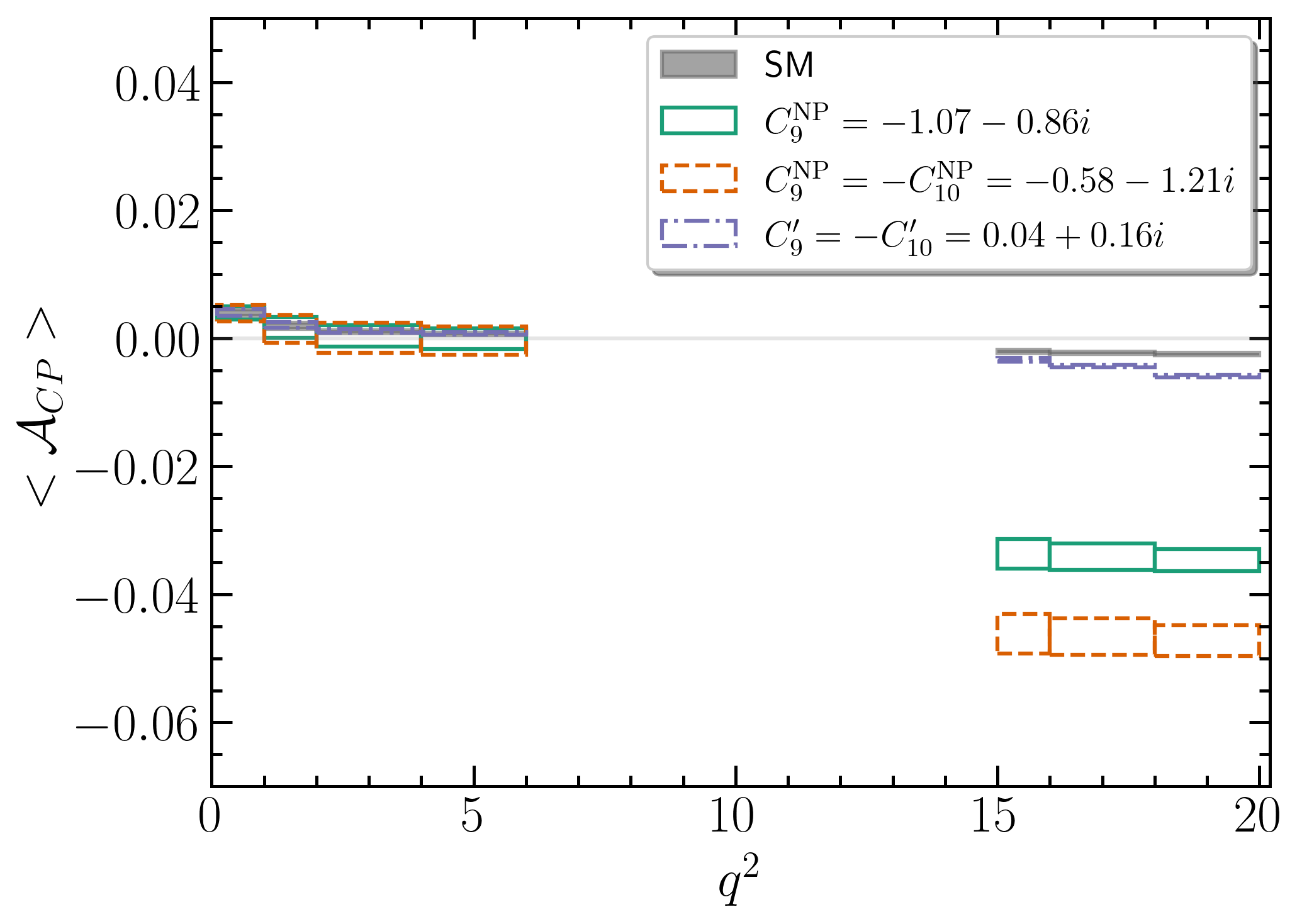}
\caption{Predictions for direct CP asymmetry in decay rate ($\A_{CP}$) of $\LbLmm$.
In the left plot, the theoretical uncertainties are shown only for the SM case
as dark grey band. In the right plot, the width of rectangle boxes denotes $q^2$-bin size, and the height of the boxes shows prediction of the observable together with corresponding ($1\sigma$) uncertainties. The same style is used in the rest of  figures of this paper.}
\label{fig: acp}
\end{figure}

In Fig.~\ref{fig: acp}, we show the ${\cal A}_{CP}$
as function of $q^2$ (left plot) in the SM and NP scenarios.
The corresponding binned predictions are shown in the plot to the right.
We find that in the SM the ${\cal A}_{CP}$ is, as expected,
very small and is $\sim \ord(10^{-3})$. In the NP case, we note the following: 
\begin{itemize}
	\item ${\cal A}_{CP}$ is sensitive to left-handed NP at large $q^2$ as seen
	from Fig.~\ref{fig: acp} where NP cases I and II show a large deviation from the SM.
	We find that in case II, which has WCs with larger imaginary part, ${\cal A}_{CP}$
	can be up to $5\%$.
	\item We find that ${\cal A}_{CP}$ is not sensitive to right-handed currents (NP case III).
	\item At low $q^2$, we find  ${\cal A}_{CP}$ is not sensitive to any of NP scenarios. 
\end{itemize}

\subsection{CP asymmetry in longitudinal polarization}
We define the CP asymmetry in longitudinal polarization fraction as the difference between
longitudinal polarization of $\LbLnp$ (denoted as $F_L(q^2)$) and
its CP-conjugate decay (denoted as $\bar F_L(q^2)$), normalized by the sum of corresponding decay rates,
\begin{align}\label{eq: a_FL}
	{\cal A}_{F_L} = \frac{F_L(q^2)- \bar F_L(q^2) }
						  {d \Gamma/dq^2 + d \bar\Gamma/dq^2},
\end{align}
where $F_L(q^2)$ and $\bar F_L(q^2)$ in terms of angular coefficients
are given as,
\begin{align}\label{eq: FL}
	F_L(q^2) = 2 K_{1ss} - K_{1cc},\quad 
	\bar F_L(q^2) = 2 \bar K_{1ss} - \bar K_{1cc}\,.
\end{align}

\begin{figure}[h]
\center
\includegraphics[width=0.49\textwidth]{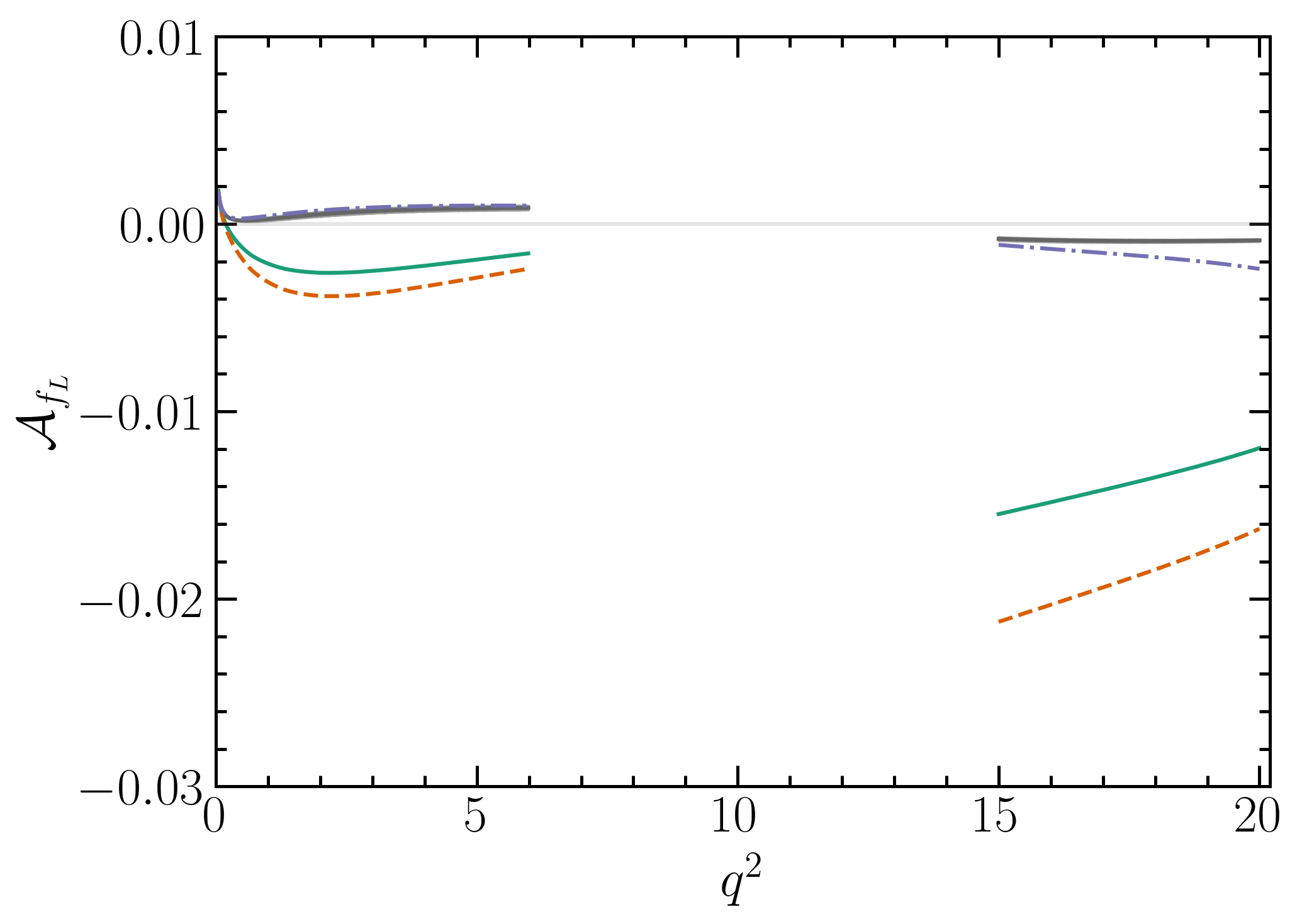}
\includegraphics[width=0.49\textwidth]{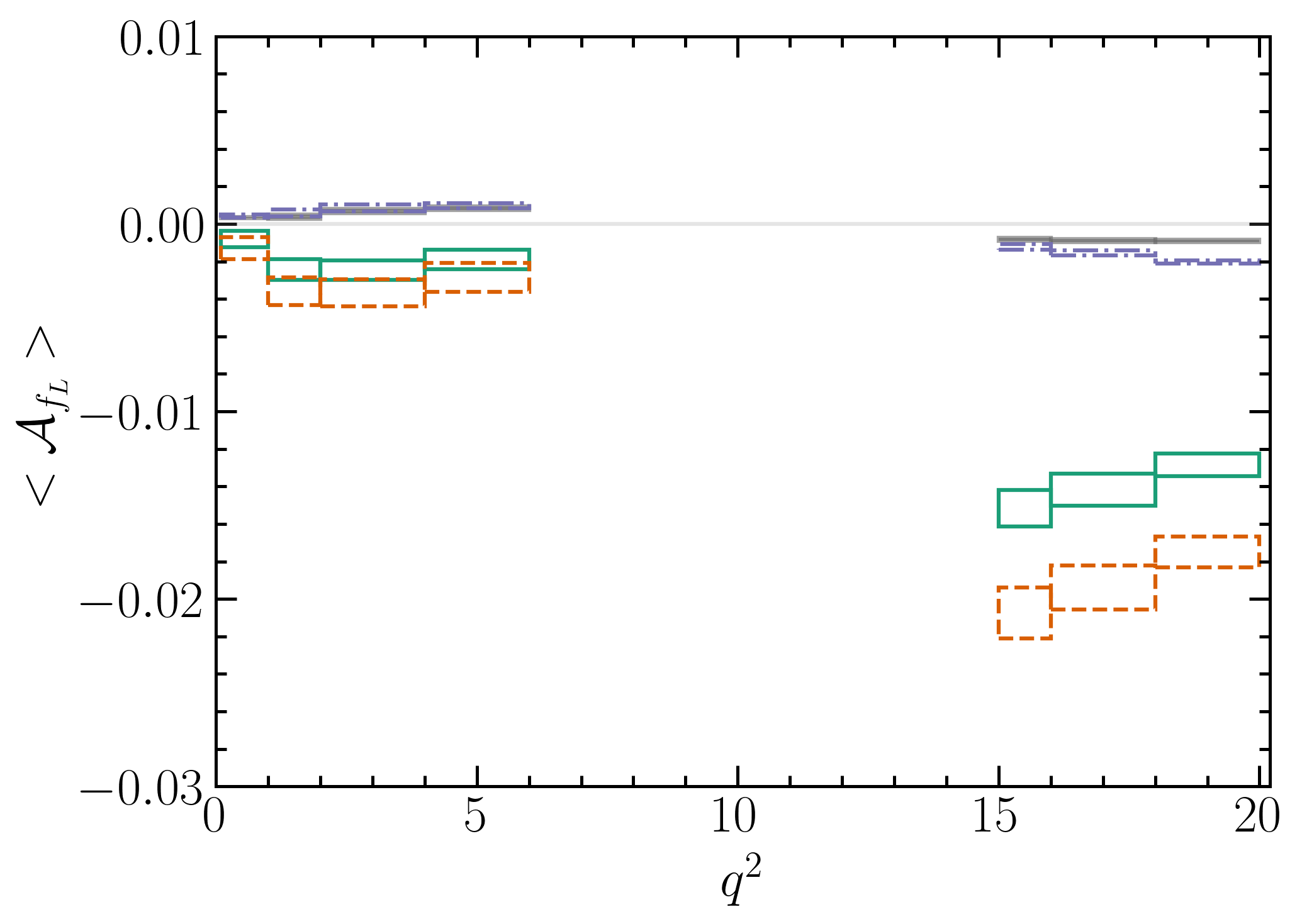}
\caption{Predictions for CP asymmetry in longitudinal polarization 
of $\LbLmm$.}
\label{fig: a_FL}
\end{figure}

Since angular coefficients appearing in ${\cal A}_{f_L}$ are the same as in ${\cal A}_{CP}$,
one expects ${\cal A}_{f_L}$ to be sensitive to NP. In Fig.~\ref{fig: a_FL},
we show results for asymmetry ${\cal A}_{f_L}$, and indeed we find that observable
is sensitive to left-handed NP (NP cases I and II) while very weakly sensitive to
right-handed NP (NP case III). The NP effects are higher in the large $q^2$ region and
 ${\cal A}_{f_L}$ can be $\sim 2\%$. In the low $q^2$ region, 
${\cal A}_{f_L}$ remains well below $1\%$.

\subsection{CP asymmetry in forward-backward asymmetries}
\begin{figure}[]
\center
\includegraphics[width=0.49\textwidth]{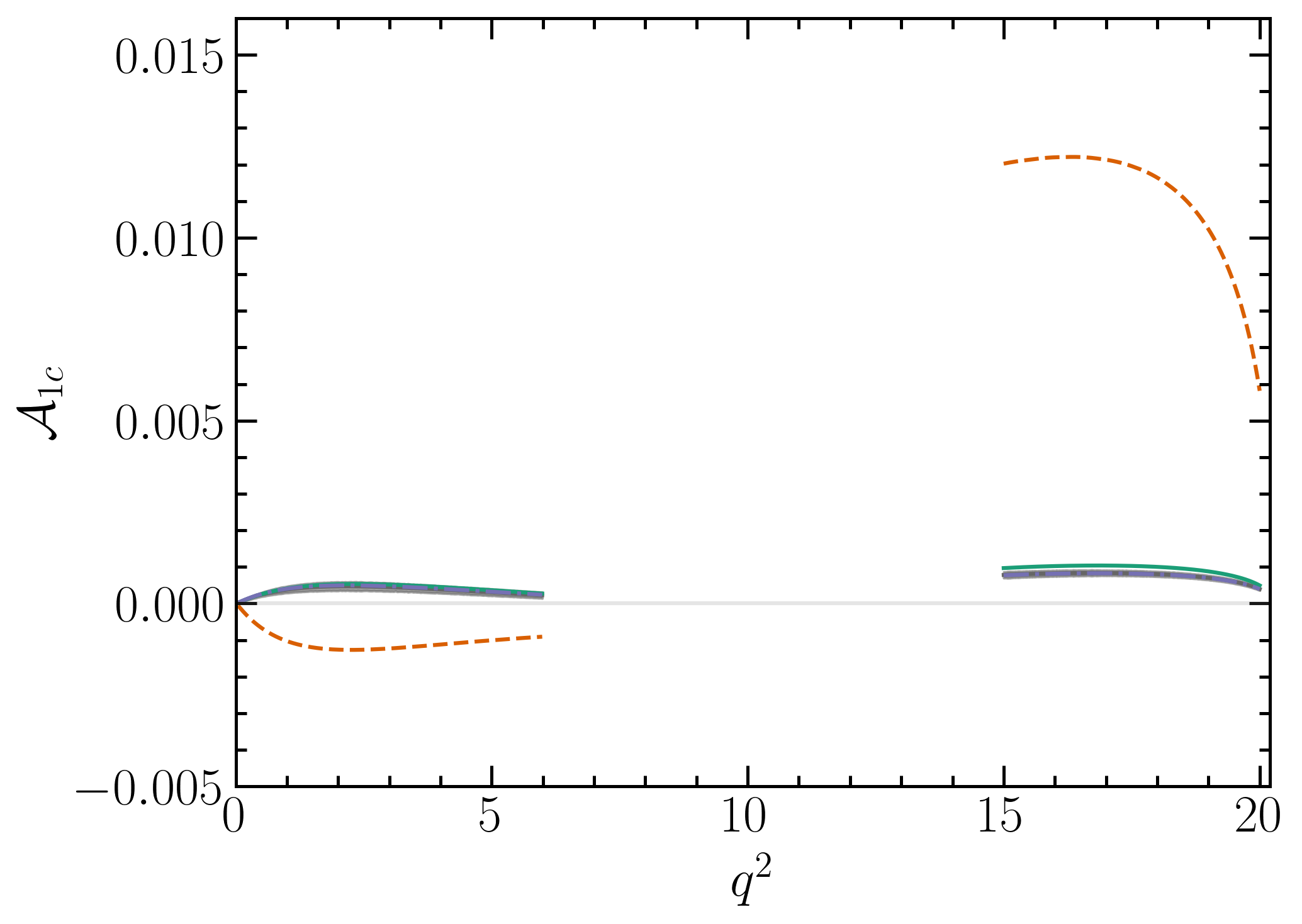}
\includegraphics[width=0.49\textwidth]{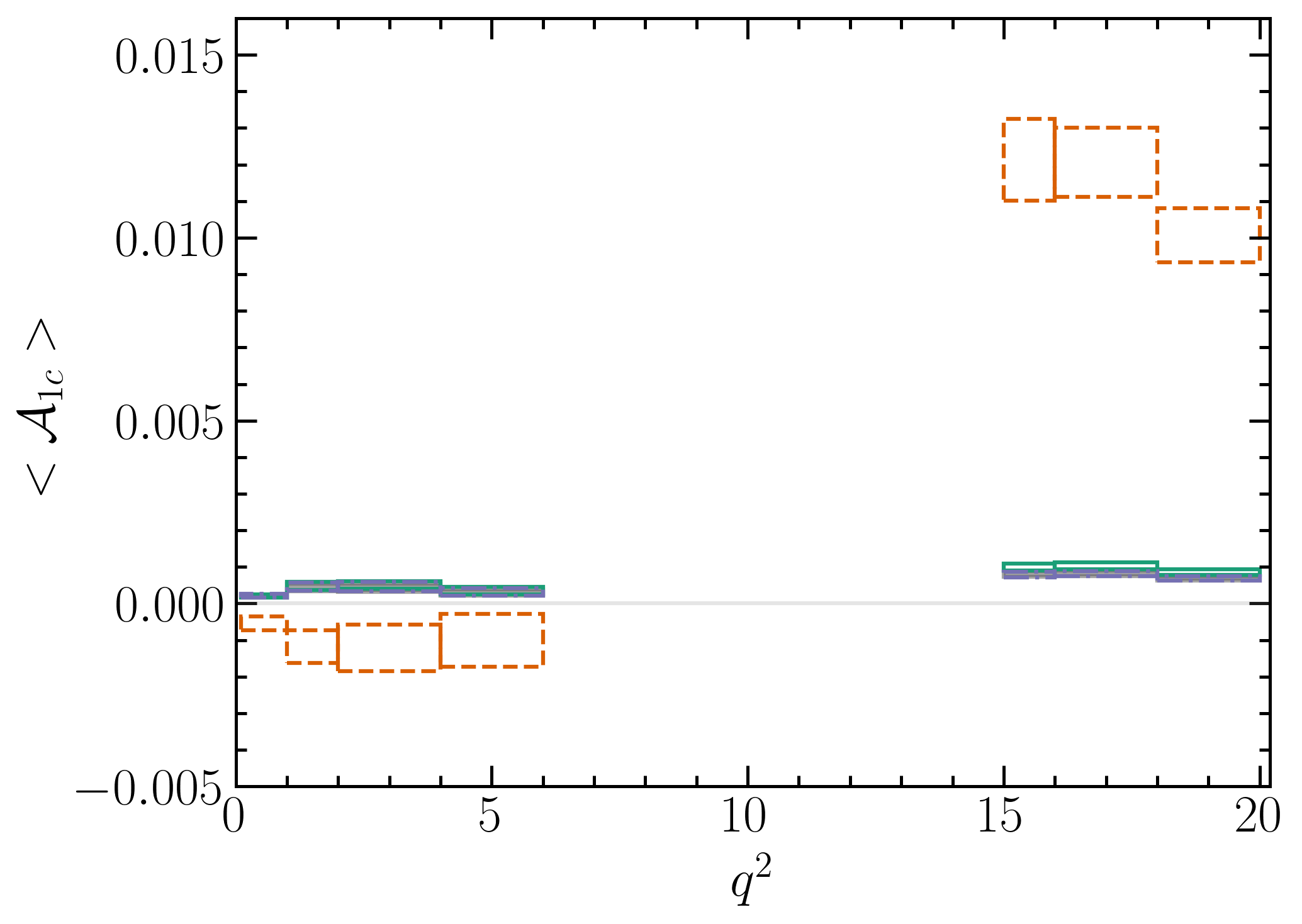}
\includegraphics[width=0.49\textwidth]{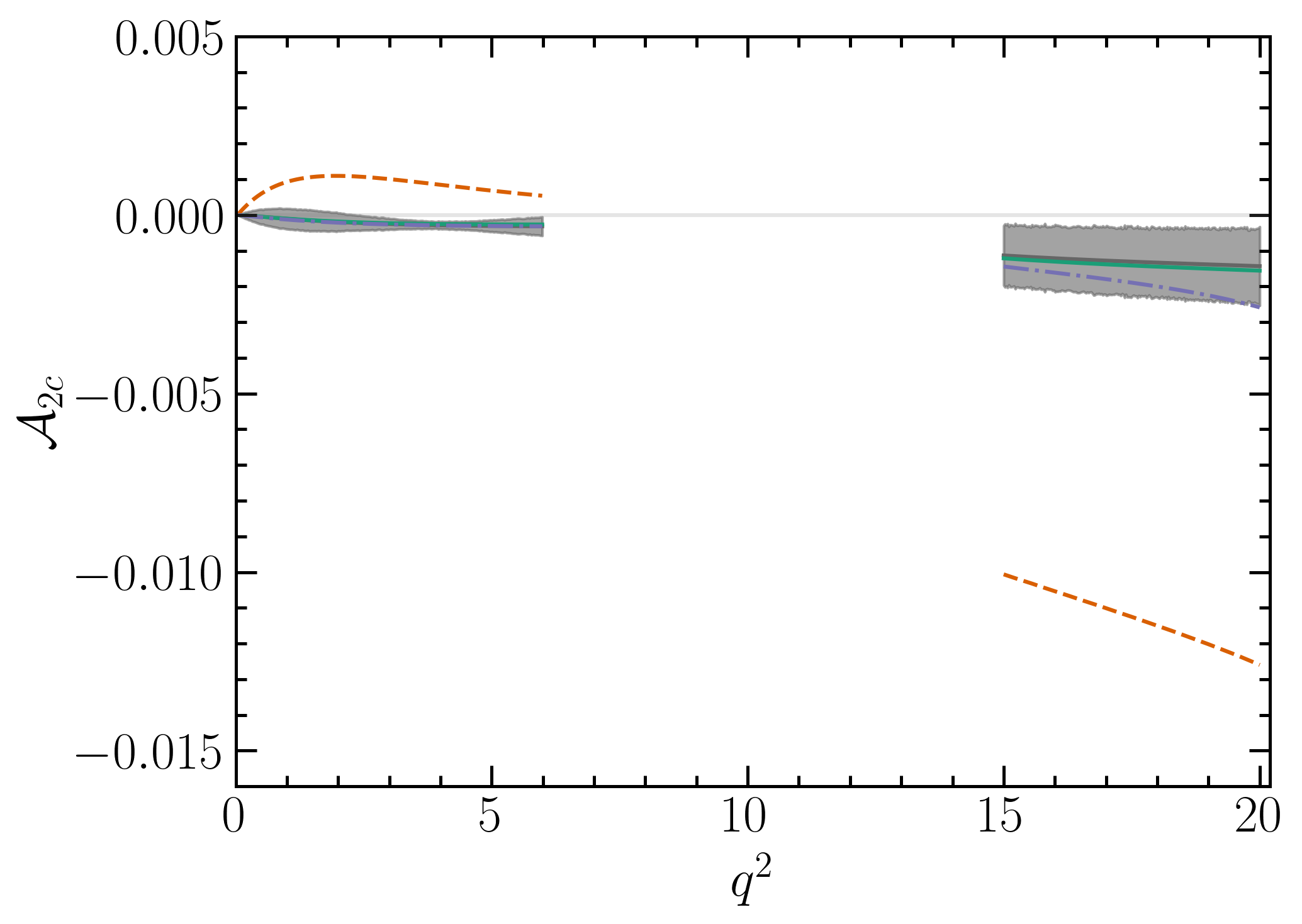}
\includegraphics[width=0.49\textwidth]{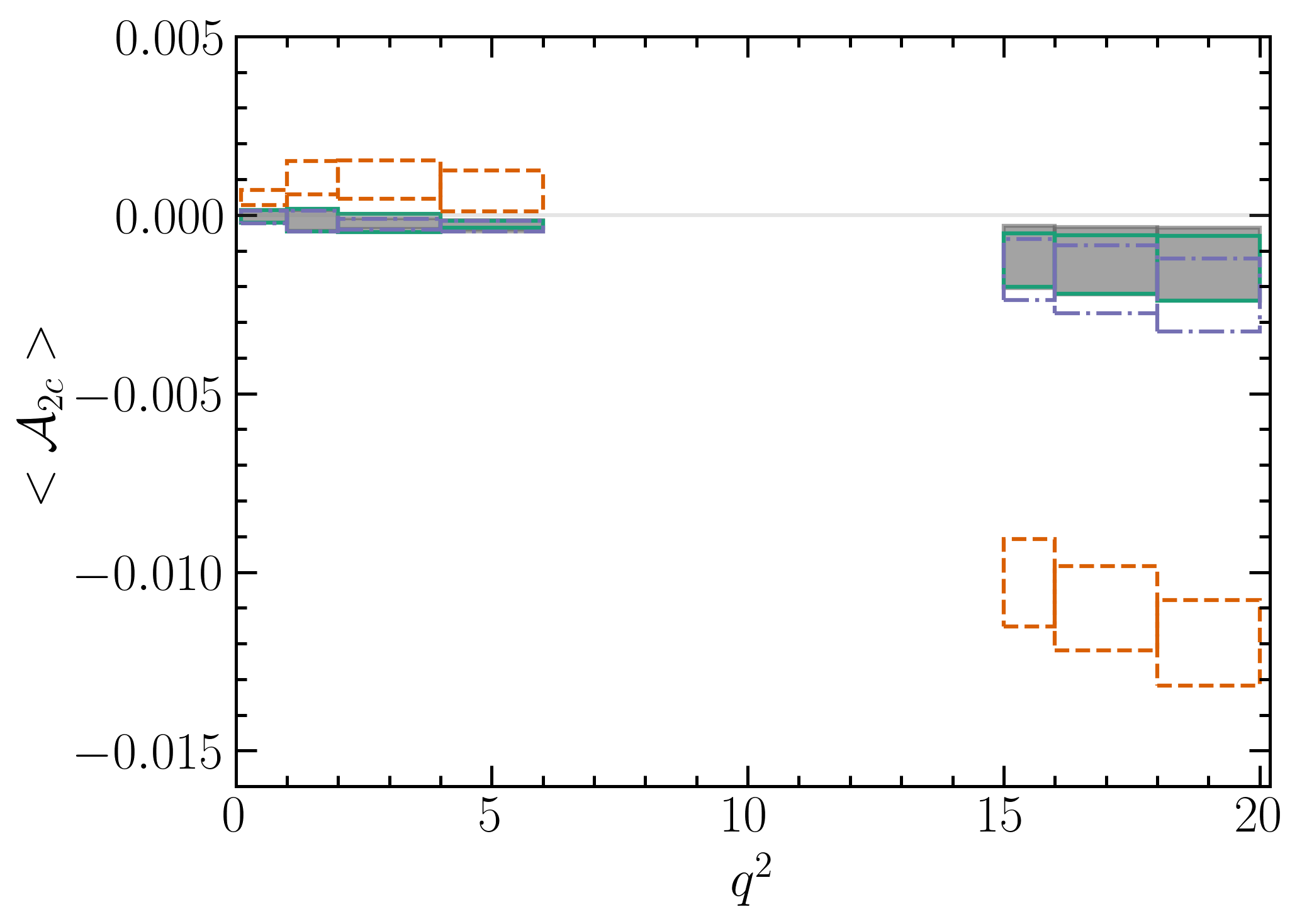}
\includegraphics[width=0.49\textwidth]{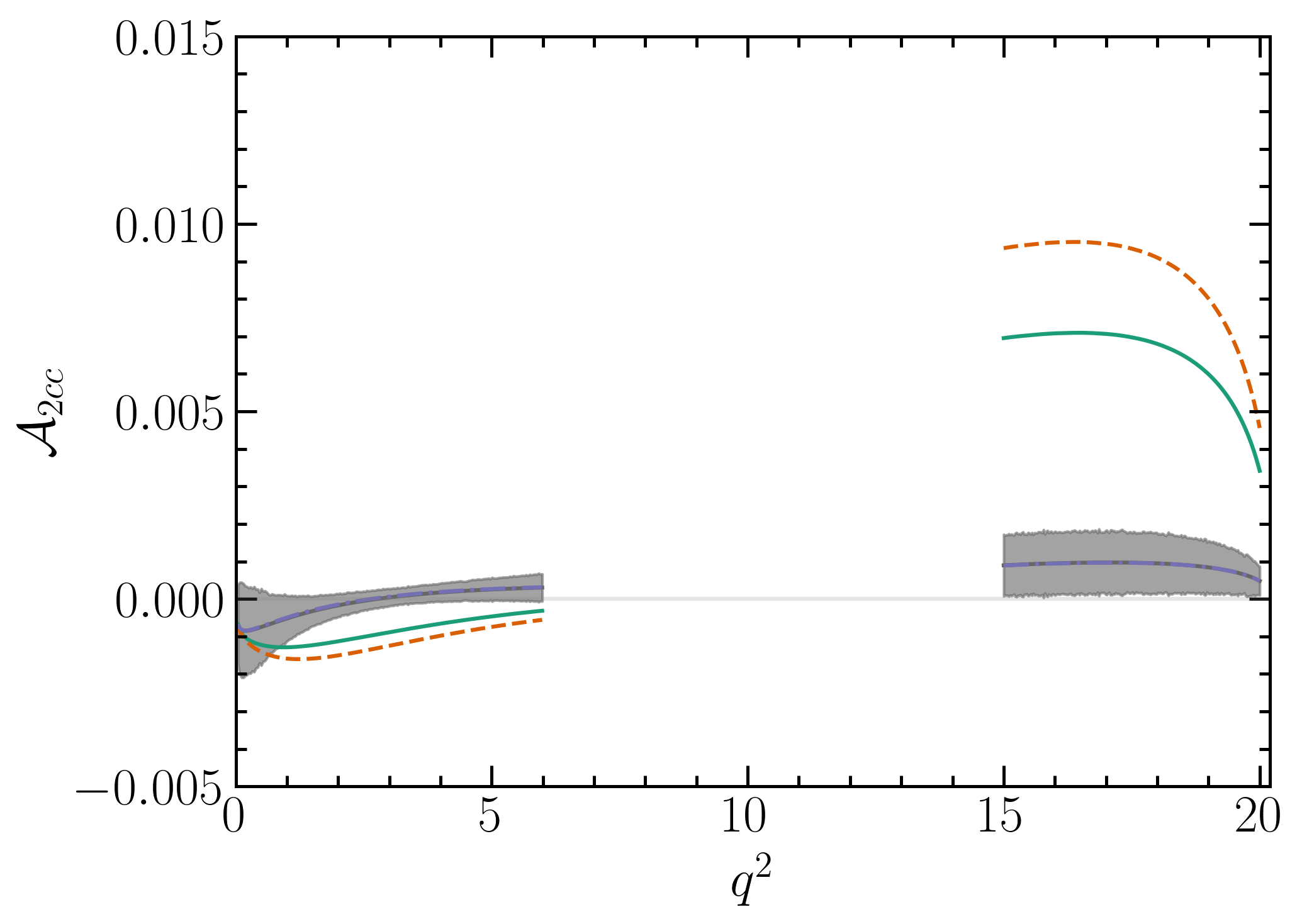}
\includegraphics[width=0.49\textwidth]{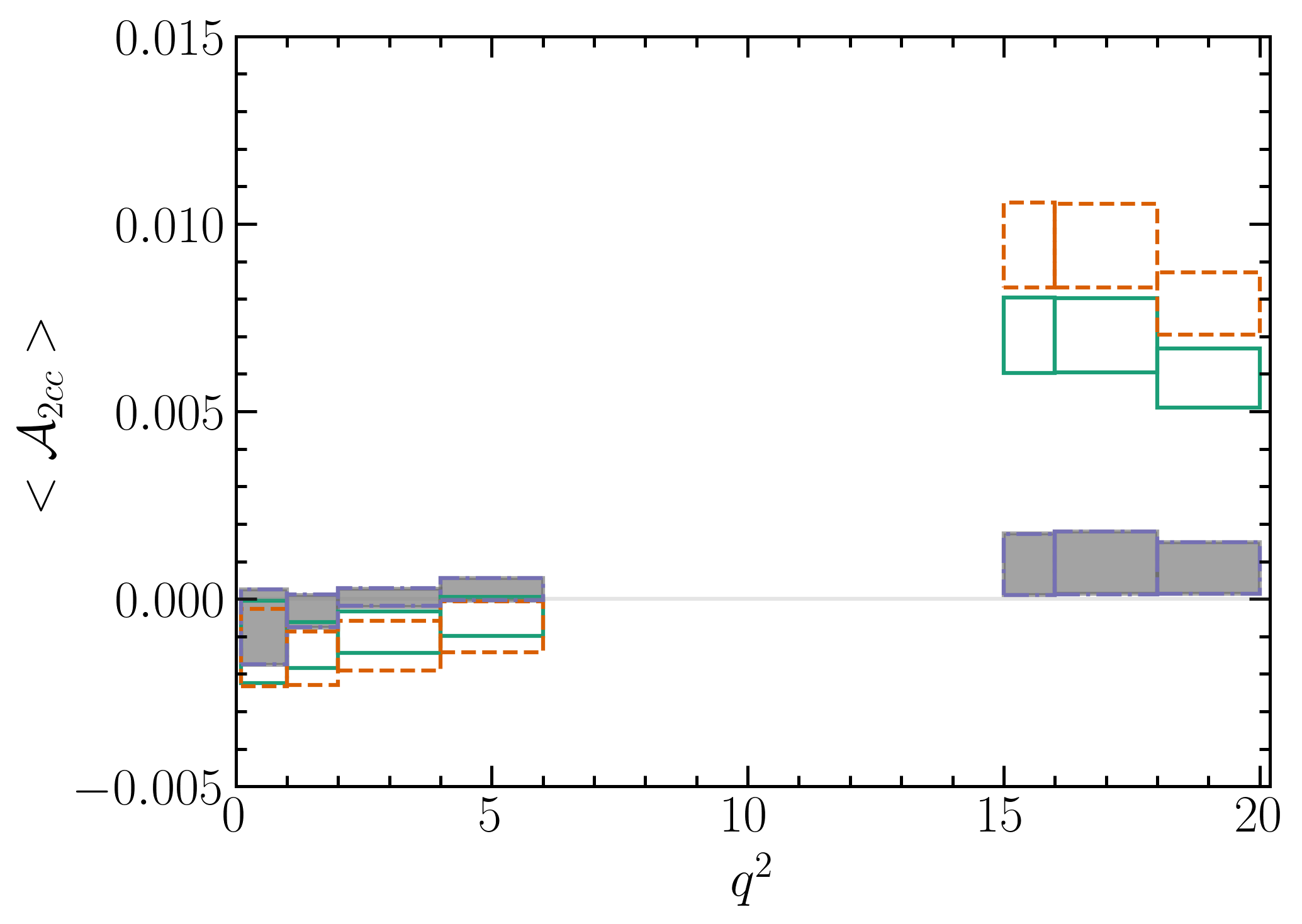}
\includegraphics[width=0.49\textwidth]{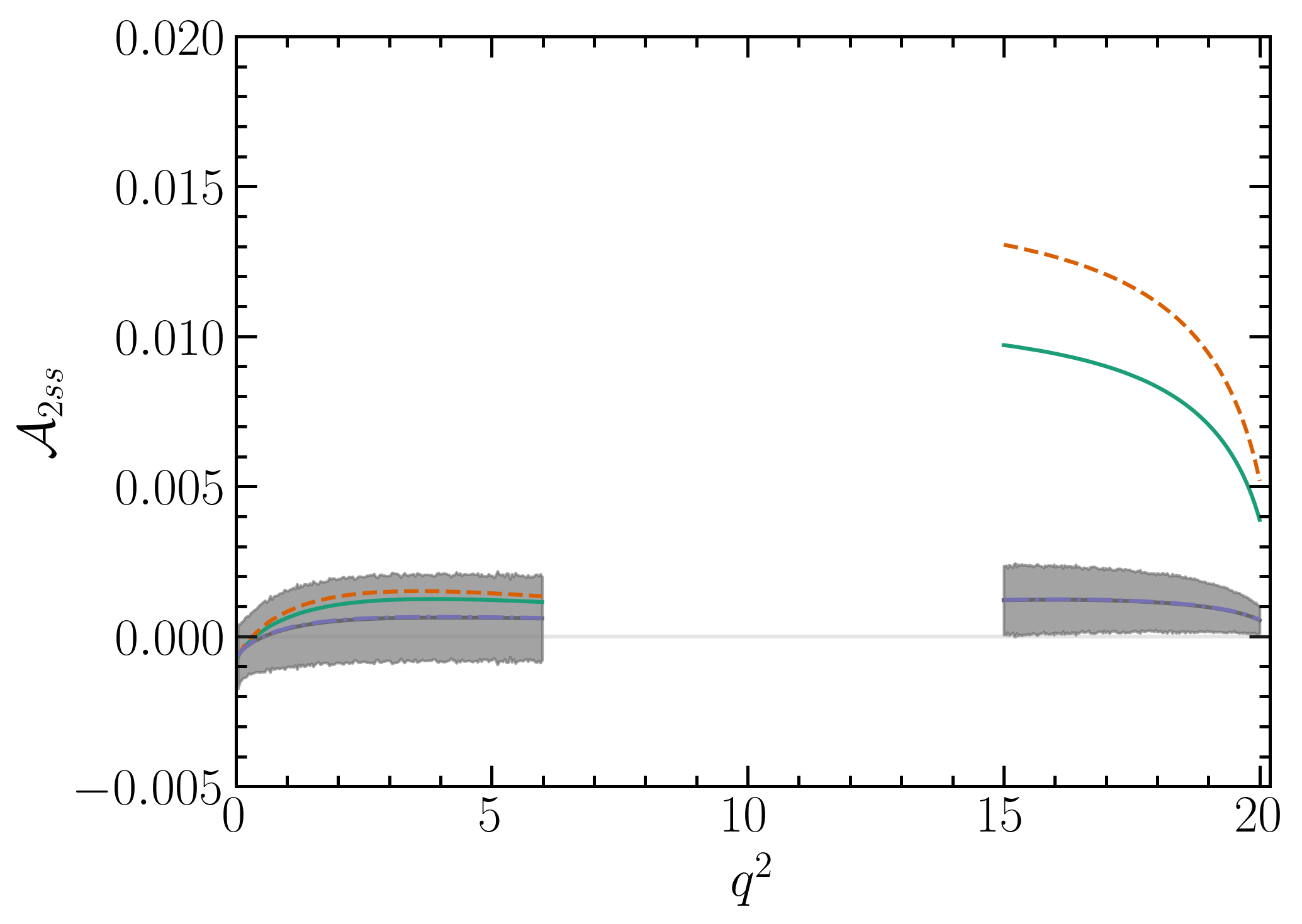}
\includegraphics[width=0.49\textwidth]{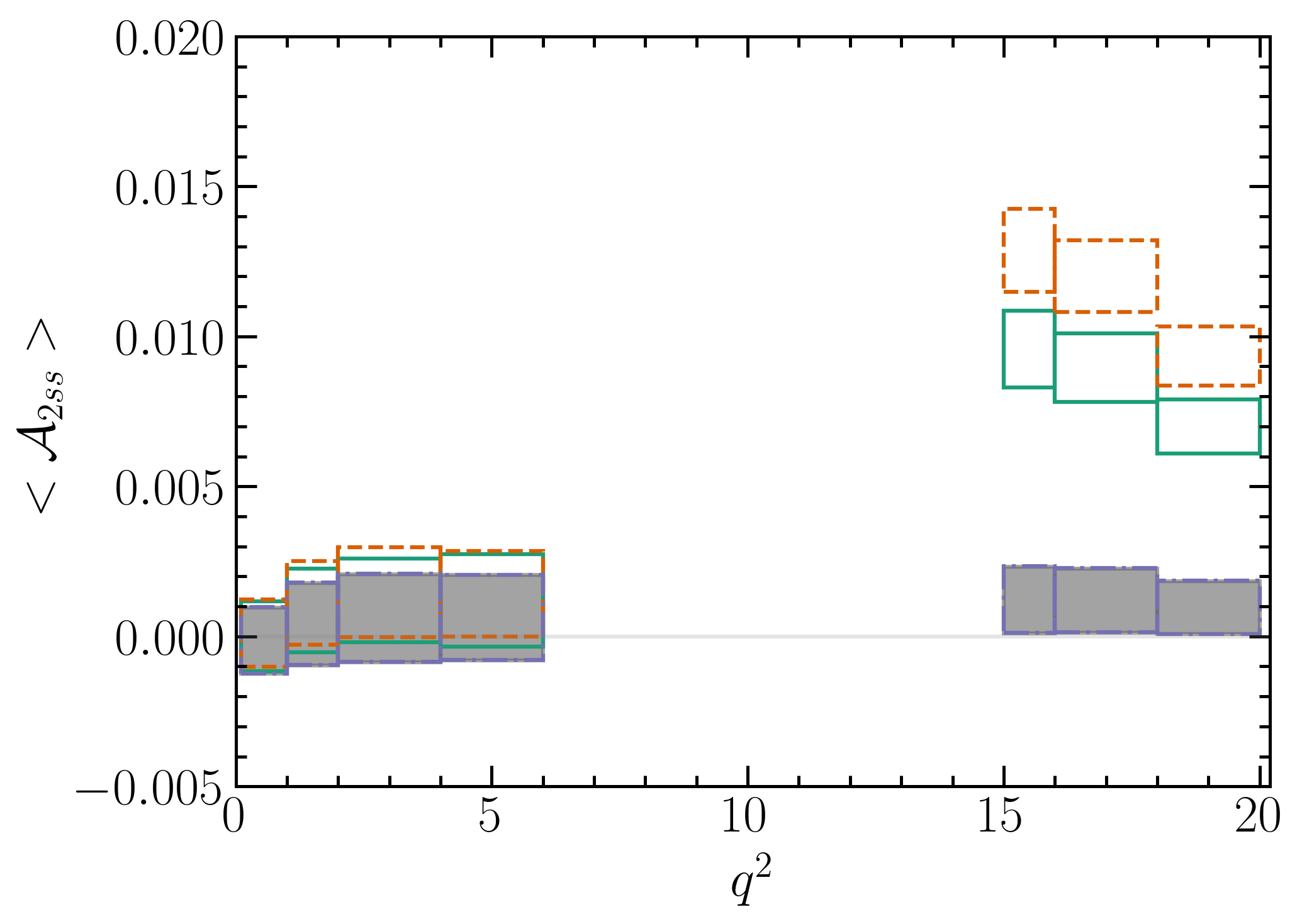}
\caption{Predictions for CP asymmetries ${\cal A}_{1c}$, ${\cal A}_{2c}$,
${\cal A}_{2cc}$, and ${\cal A}_{2ss}$.}
\label{fig: a_12}
\end{figure}

The decay $\LbLnp$ offers three types of forward-backward (FB) asymmetries \cite{Boer:2014kda}:
 FB asymmetry ($a_{\rm FB}^\ell$) with respect to leptonic angle
 $\theta_\ell$, FB asymmetry ($a_{\rm FB}^\Lambda$) with respect to hadronic angle
 $\theta_\Lambda$, and FB asymmetry ($a_{\rm FB}^{\ell \Lambda }$) with respect to the combination of
 $\theta_\ell$ and $\theta_\Lambda$, respectively.
 In terms of angular coefficients, these are given as,
	\begin{align}
		a^\ell_{\rm FB} = \frac{3}{2} K_{1c}, \quad 
		a^{\Lambda}_{\rm FB} = \frac{1}{2} (2K_{2ss} + K_{2cc}), \quad 
		a^{\ell \Lambda}_{\rm FB} = \frac{3}{4}K_{2c}\,.
	\end{align}
One can then take the difference between the measurement of the FB asymmetry in the
decay $\LbLnp$ and its CP-conjugate to define new CP asymmetries. These CP asymmetries
can be accessed by measuring the CP asymmetries associated with angular coefficients
$K_{1c}$, $K_{2c}$, $K_{2ss}$, and $K_{2cc}$.
To this end, we define the following four CP asymmetries:
\begin{align}\label{eq: a_1c}
{\cal A}_{1c} &= \frac{K_{1c}(q^2)-\bar K_{1c}(q^2)}{d \Gamma/dq^2 + d \bar\Gamma/dq^2},\\
{\cal A}_{j} &= \frac{K_{j}(q^2)+\bar K_{j}(q^2)}{d \Gamma/dq^2 + d \bar\Gamma/dq^2},	\quad \text{for}\ j = 2c, 2ss, 2cc.
\label{eq: a_2c}
\end{align}
${\cal A}_{1c}$ and ${\cal A}_{2c}$  are equivalent to CP asymmetries in
$a^\ell_{\rm FB}$ and $a^{\ell \Lambda}_{\rm FB}$ up to a normalization constant,
while CP asymmetry in $a^{\Lambda}_{\rm FB}$ can be determined from combined measurements
of  ${\cal A}_{2ss}$ and ${\cal A}_{2cc}$. Also, note that ${\cal A}_{1c}$ involves the 
difference of $K_{1c}$ and $\bar K_{1c}$, while the other CP
asymmetries involves the sum of corresponding coefficients.
This is because the angular coefficients in 
Eqs.~\eqref{eq: coeff-K-first}--\eqref{eq: coeff-K-end} are proportional to the decay parameter
$\alpha_\Lambda$
(or $\bar \alpha_\Lambda$ in the case of CP-conjugate mode). As discussed in the previous
section, $\alpha_\Lambda \simeq - \bar\alpha_\Lambda$ experimentally, therefore it is the
combination $K_j + \bar K_j$ ($j = 2c, 2ss, 2cc, 3s, 3sc, 4s, 4sc$) that vanishes in the 
case of purely real WCs.

In Fig.~\ref{fig: a_12}, we show results for 
${\cal A}_{1c}$, ${\cal A}_{2c}$, ${\cal A}_{2cc}$, and ${\cal A}_{2ss}$.
We note the following:
\begin{itemize}
	\item CP asymmetry $\A_{1c}$ is sensitive to $C_{10}$, while no sensitivity to
	vector and right-handed current is found.
	We find that in NP case II,  which has nonzero $C_{10}^{\rm NP}$,
	the value of $\A_{1c}$ can be at $\sim 1.3\%$ level in the large $q^2$ region.
	In other NP cases considered,  $\A_{1c}$ remain indistinguishable from the SM.
	\item For the asymmetry ${\cal A}_{2c}$, we observe similar NP sensitivity as in $\A_{1c}$
	except that the sign of the asymmetry is opposite to that of $\A_{1c}$. We also note that
	${\cal A}_{2c}$ is largest ($\sim 1.3\%$) at the kinematic end point
	$q^2\sim 20~{\rm GeV}^2$.
	\item In contrast, the asymmetries  ${\cal A}_{2ss}$ and ${\cal A}_{2ss}$
	are sensitive to $C_9$, while no sensitivity to
	axial vector and right-handed currents is found.
	Both asymmetries are found to be $\sim 1\%$ in large $q^2$ region,
	with ${\cal A}_{2ss}$ being somewhat slightly larger. 
\end{itemize}

\subsection{More CP asymmetries}
One can define four more CP asymmetries associated with angular coefficients
$K_{3s}$,  $K_{3sc}$, $K_{4s}$, and  $K_{4sc}$, respectively,
\begin{align}
	{\cal A}_{j} &= \frac{K_{j}(q^2)+\bar K_{j}(q^2)}{d \Gamma/dq^2 + d \bar\Gamma/dq^2},	\quad \text{for}\ j = 3s, 3sc, 4s, 4sc.
\end{align}

\begin{figure}[]
\center
\includegraphics[width=0.49\textwidth]{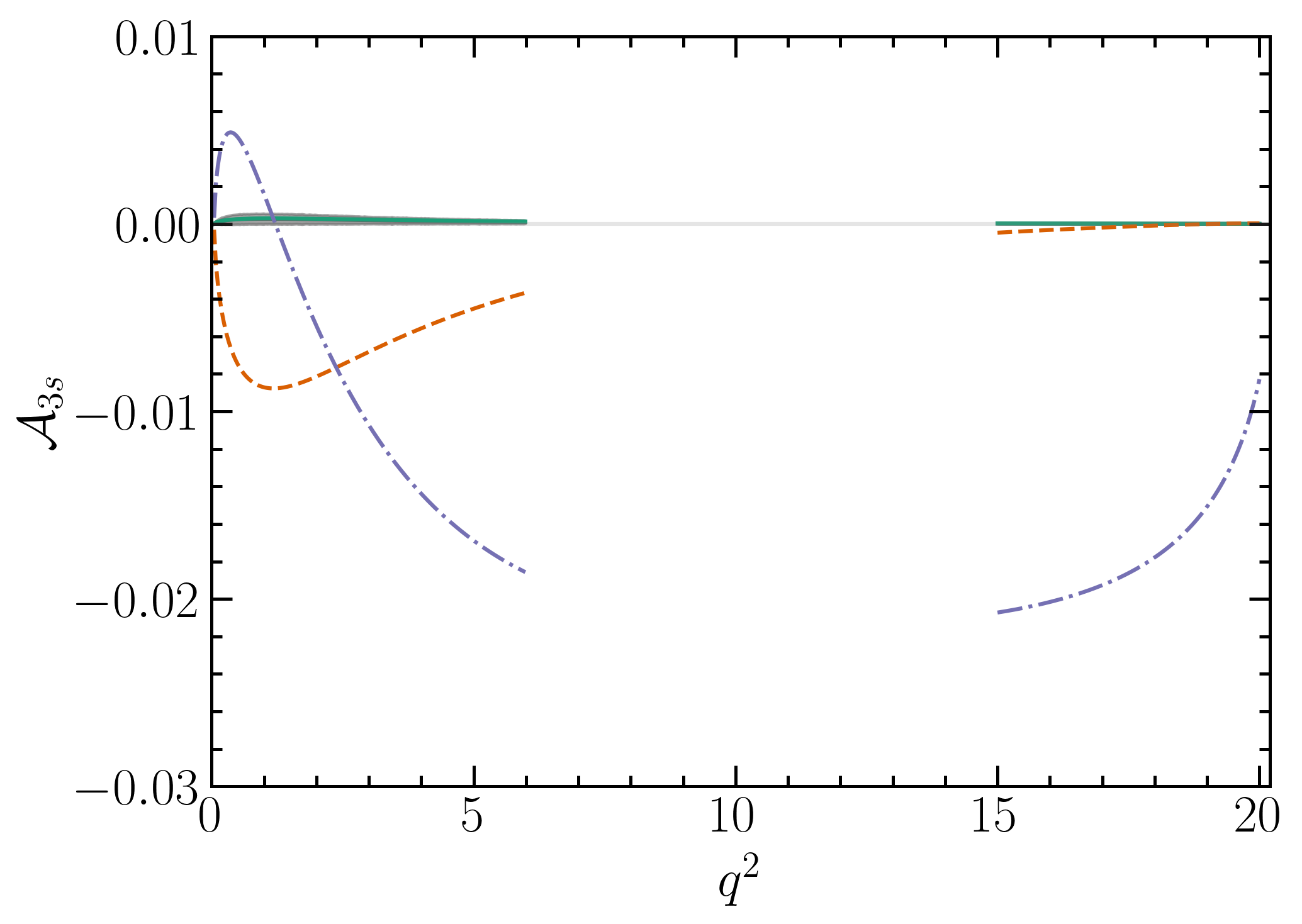}
\includegraphics[width=0.49\textwidth]{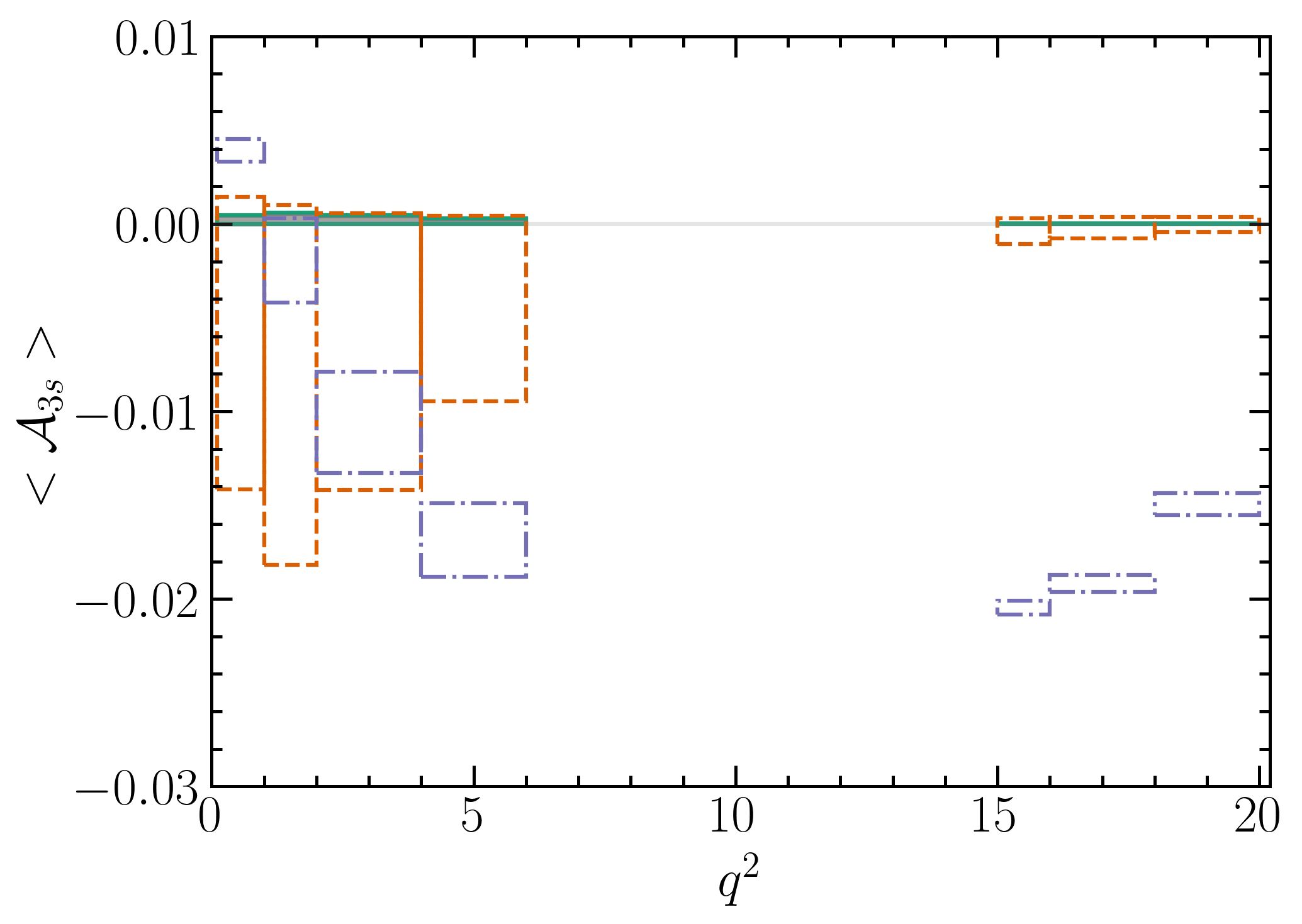}
\includegraphics[width=0.49\textwidth]{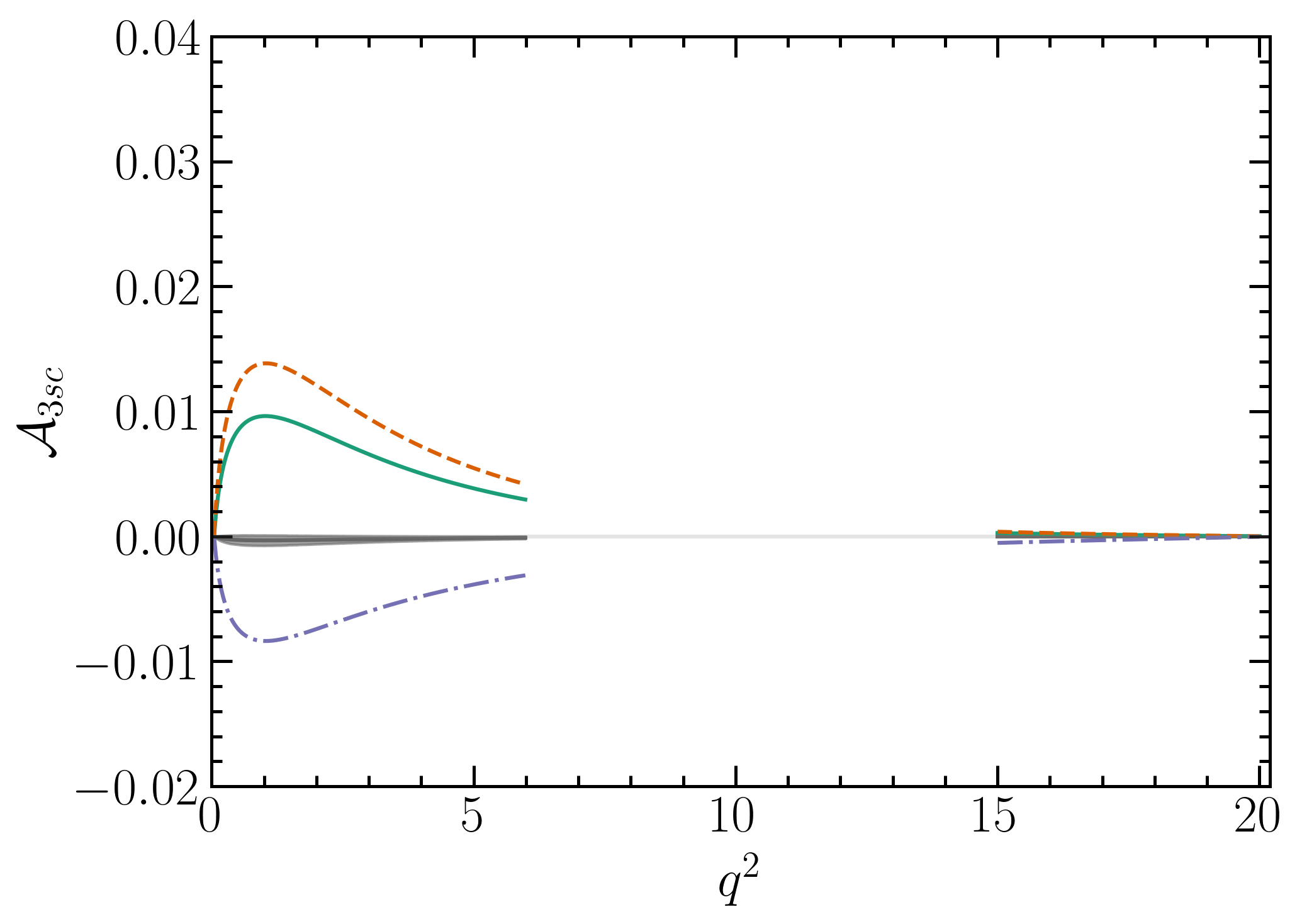}
\includegraphics[width=0.49\textwidth]{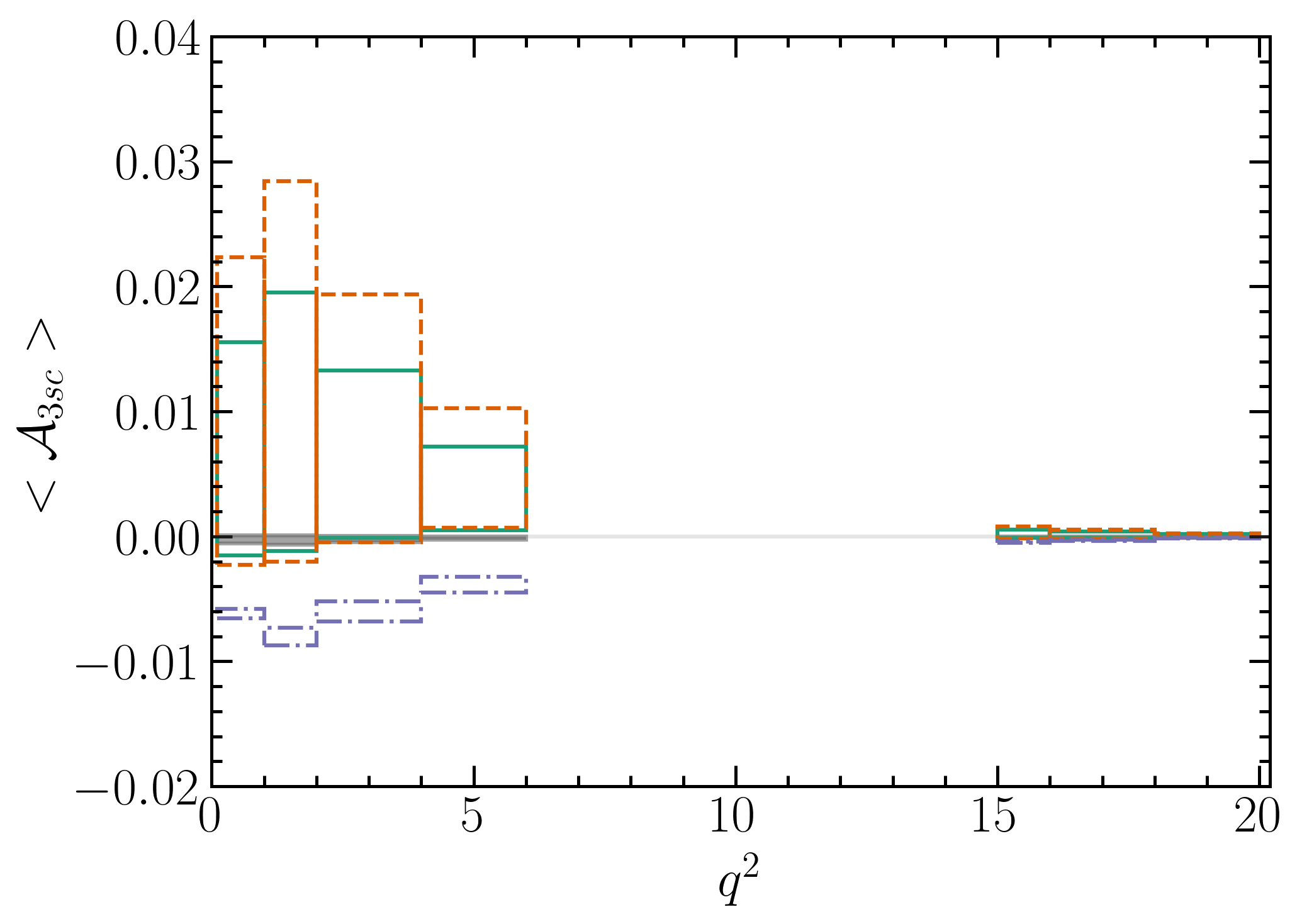}
\includegraphics[width=0.49\textwidth]{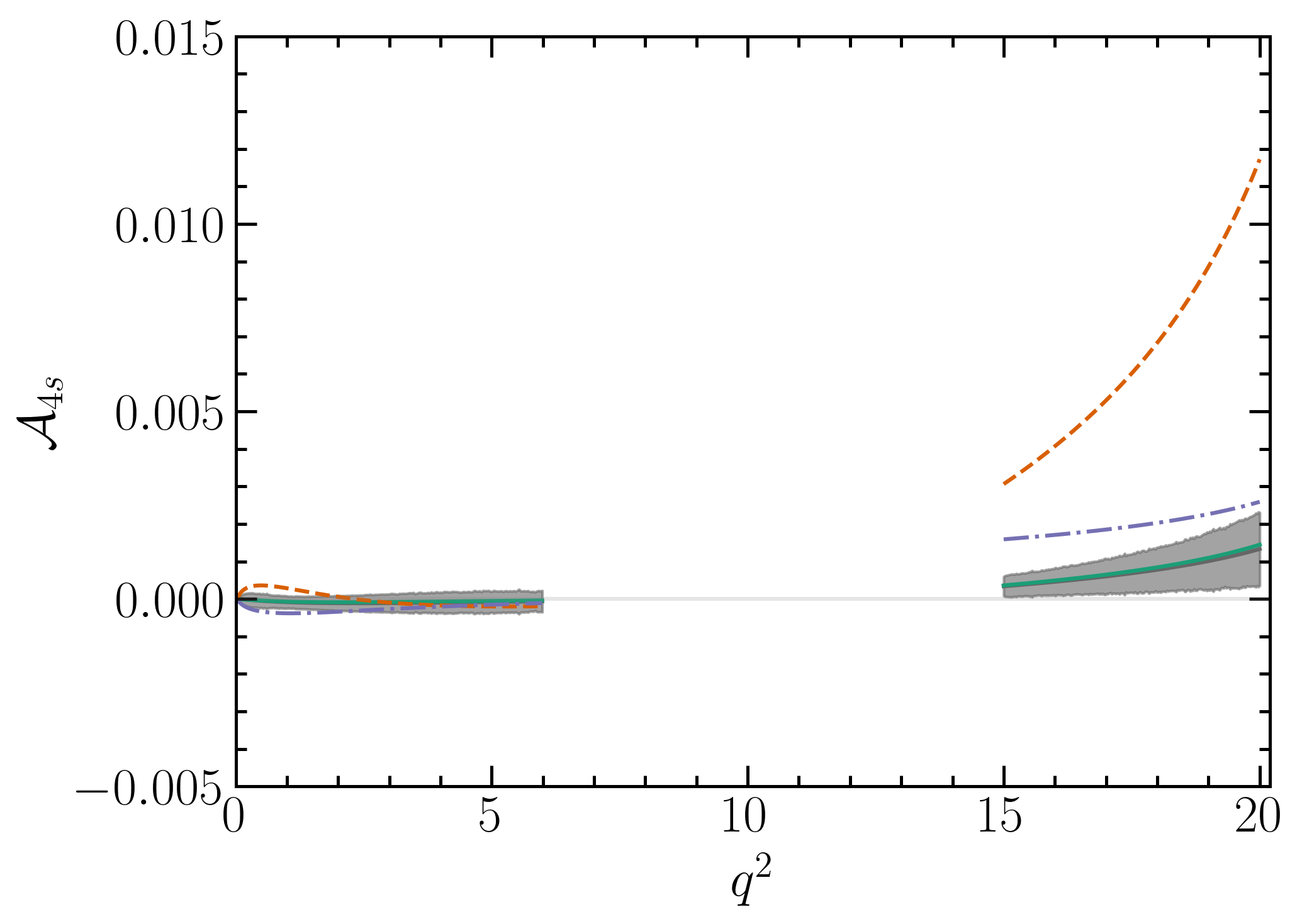}
\includegraphics[width=0.49\textwidth]{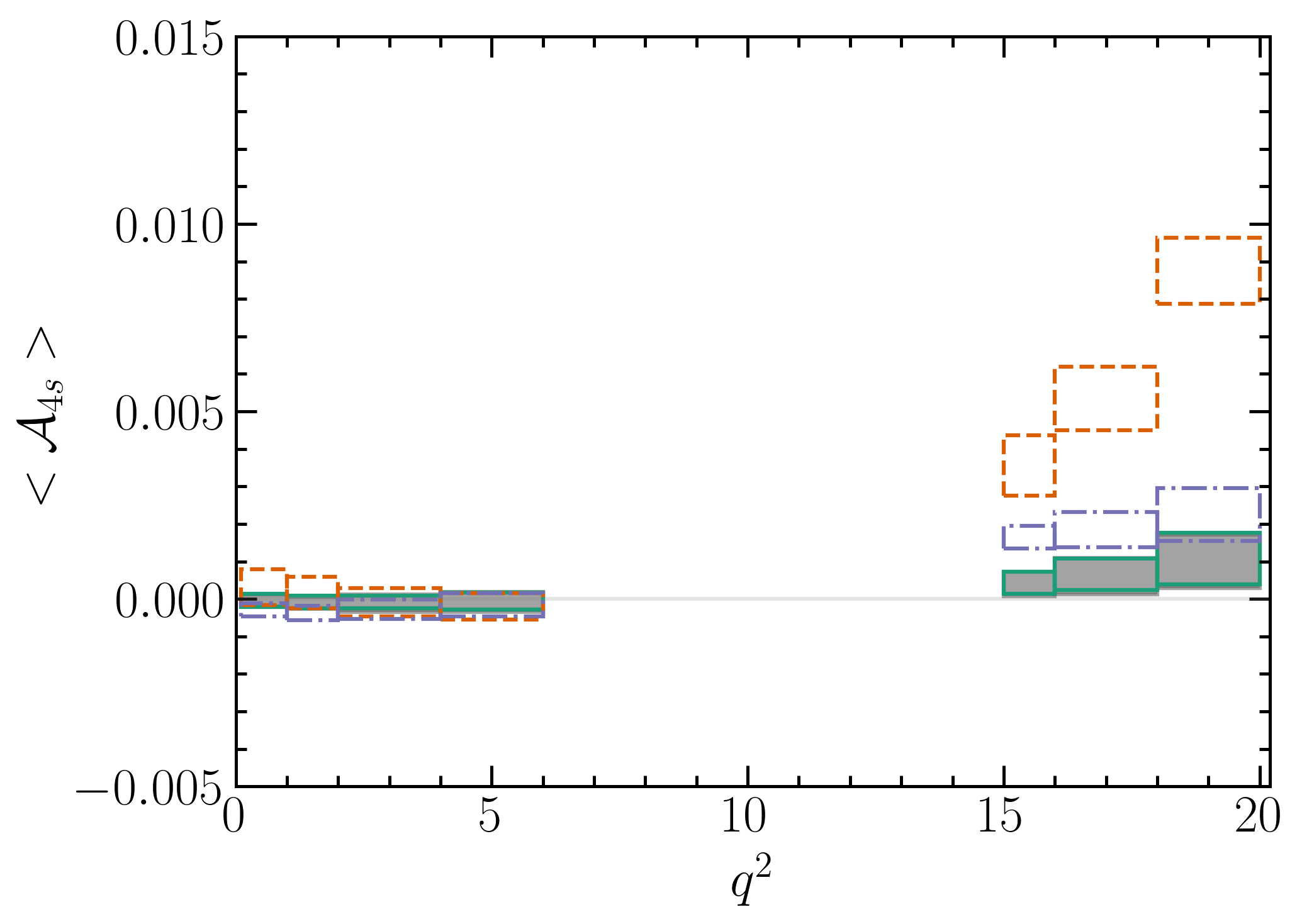}
\includegraphics[width=0.49\textwidth]{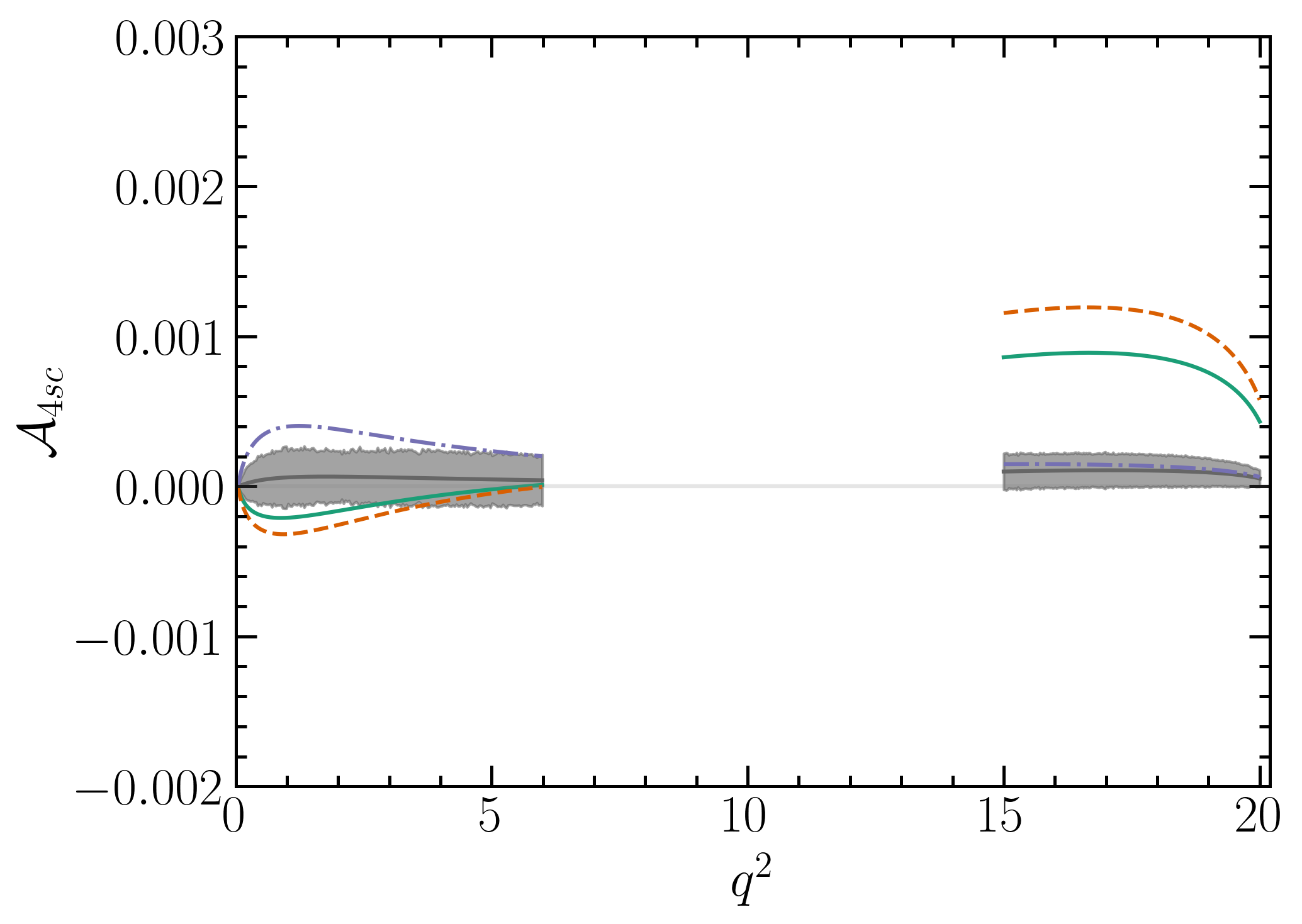}
\includegraphics[width=0.49\textwidth]{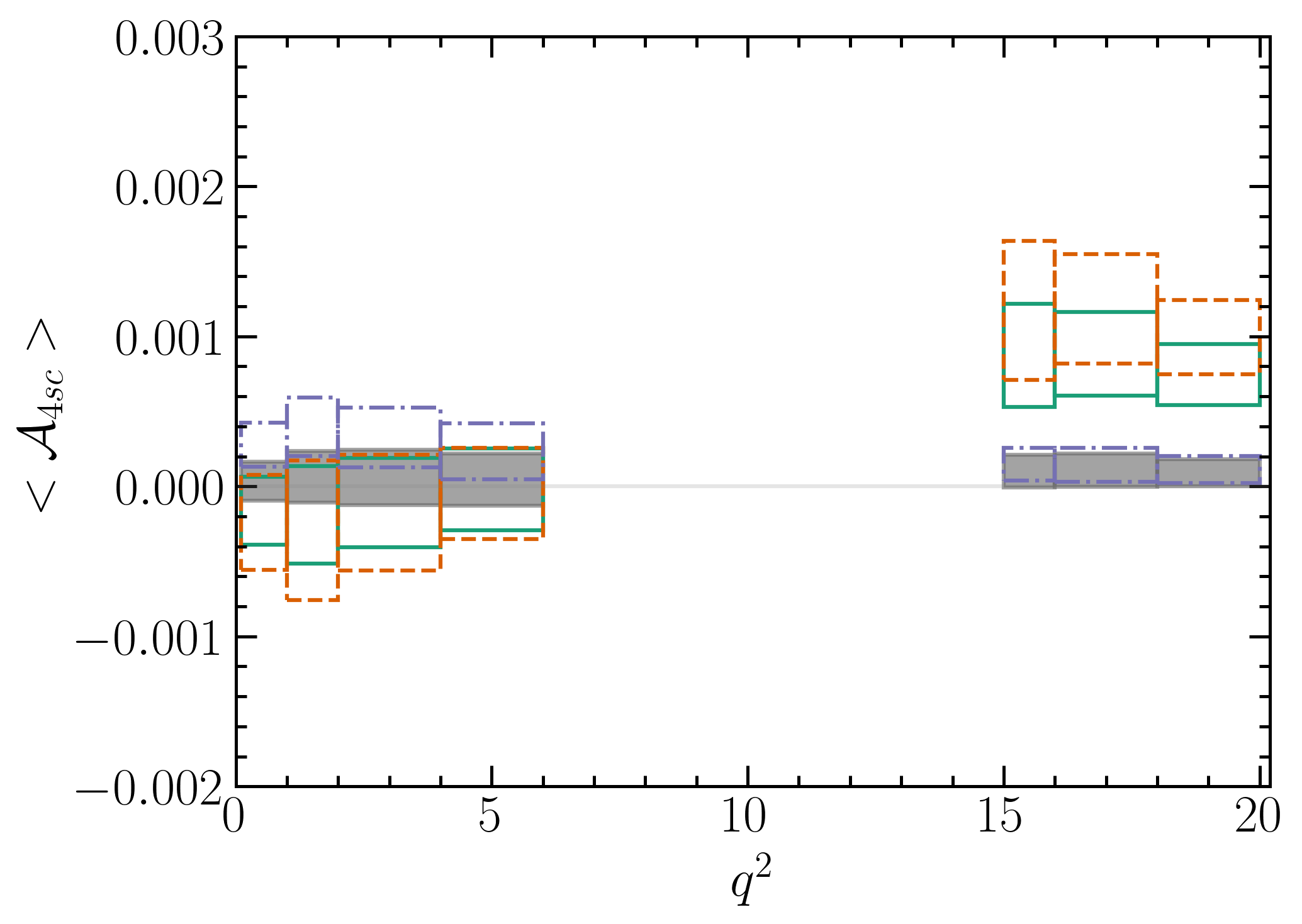}
\caption{Predictions for CP asymmetries ${\cal A}_{3s}$, ${\cal A}_{3sc}$,
${\cal A}_{4s}$, and  ${\cal A}_{4sc}$.}
\label{fig: a_34}
\end{figure}

In Fig.~\ref{fig: a_34} we show the results for these CP asymmetries and make
the following observations:

\begin{itemize}
	\item We find ${\cal A}_{3s}$ to be   sensitive to right-handed currents
	(NP case III) in both low and large $q^2$ regions, where it can be up to $\sim 2\%$. 
	It is also mildly sensitive to $C_{10}$ in the low $q^2$ region,
	but has large theoretical uncertainties, as can be seen from low $q^2$ bins.
	Another point worth mentioning is that there is zero-crossing in low $q^2$ region
	of ${\cal A}_{3s}$  for right-handed NP scenario
	(case III), while no such behavior is seen in left-handed NP scenarios.
	\item In case of ${\cal A}_{3sc}$, we find the CP asymmetry to be sensitive to $C_9$
	 as well as right-handed NP case, with its size being $\sim 1-3\%$ in low $q^2$ region.
	 At large $q^2$, CP asymmetry remains negligible. 
	 \item For ${\cal A}_{4s}$, we find that the asymmetry is sensitive to
	 $C_{10}$ at large $q^2$. The CP asymmetry remains SM-like in cases with NP in $C_9$ or
	 right-handed NP. In NP case II (which has nonzero $C_{10}$), ${\cal A}_{4s}$ can be
	 $\sim 1\%$.
	 \item On the other hand, for ${\cal A}_{4sc}$ we find that the observable
	 is most sensitive to $C_9$ (NP cases I and II), but the asymmetry remains very small, at $\mathcal{O}(10^{-3})$.
\end{itemize}

\section{Conclusion} \label{sec: conclusion}
The experimental data on the $b\to s \ell^+\ell^-$ transitions hints towards the presence of NP that is lepton flavor universal in nature. However, the CP properties of the possible underlying NP is unknown. One concrete way to answer this question is
provided by the measurements of CP-violating asymmetries associated with the
$b\to s \ell^+\ell^-$ transitions. In the SM, the CPV in the $b\to s \ell^+\ell^-$ transition is very small, but can be enhanced in many NP models.  Thus, measurements of sizable 
$b\to s\ell^+\ell^-$ CP asymmetries are highly motivated. With this in mind, 
in this paper,  we have investigated in a model-independent fashion
the prospects of probing CP-violating
NP in $\LbLnp$ decay. To this end,  we list all the
CP asymmetries offered by the angular distribution of an unpolarized $\Lambda_b$ 
decay $\LbLnp$. We then present their determinations in the SM and
several NP scenarios motivated by the present global fits to $b\to s\ell^+\ell^-$ data.
We find that several of the CP asymmetries, depending on their NP
sensitivity and $q^2$ region, can be enhanced to a few
percent level. More importantly, we find that the measurements
of these CP asymmetries can provide new methods to not only probe
but also potentially distinguish NP cases discussed in the paper.
Therefore, the CP asymmetries in $\LbLnp$ provide new avenues to cross-check
the SM and, in conjunction with CP asymmetries of $B\to K^\ast \ell^+\ell^-$,
can play a useful role in searching NP  in $b\to s\ell^+\ell^-$ transitions.

\vskip0.2cm
\noindent{\bf Acknowledgments} 
DD would like to thank the DST, Government of India for the INSPIRE Faculty Award (Grant No. IFA16-PH170). DD also thanks the Institute for Theoretical Physics III, University of Stuttgart for kind hospitality during various stages of the work. JD acknowledges the Council of Scientific and Industrial Research (CSIR), Government of India, for the SRF fellowship grant with File No. 09/045(1511)/ 2017-EMR-I. JD also would like to acknowledge Research Grant No. SERB/CRG/004889/SGBKC/2022/04 of the SERB, India, for partial financial support. The work of GK is supported by
NSTC 111-2639-M-002-002-ASP of Taiwan. NS would like to acknowledge support from the UK Science and Technology Facilities Council (STFC).

\appendix

\section{Prediction of CP asymmetries}\label{app: predictions}

\subsection{$\A_{CP}$ (in units of $10^{-2}$) }

\begin{table}[H]
\centering
\begin{tabular}{lllll}
\toprule
Bin & SM & Case I & Case II & Case III \\
\midrule
$[0.1, 1]$ & 0.399$\pm$0.052 & 0.401$\pm$0.104 & 0.389$\pm$0.129 & 0.405$\pm$0.052 \\
$[1, 2]$ & 0.197$\pm$0.046 & 0.170$\pm$0.165 & 0.141$\pm$0.216 & 0.209$\pm$0.047 \\
$[2, 4]$ & 0.098$\pm$0.024 & 0.038$\pm$0.168 & 0.005$\pm$0.237 & 0.112$\pm$0.030 \\
$[4, 6]$ & 0.066$\pm$0.017 & -0.010$\pm$0.162 & -0.038$\pm$0.224 & 0.079$\pm$0.020 \\
$[1.1, 6]$ & 0.098$\pm$0.021 & 0.037$\pm$0.165 & 0.006$\pm$0.223 & 0.111$\pm$0.025 \\
$[15, 16]$ & -0.201$\pm$0.021 & -3.367$\pm$0.227 & -4.614$\pm$0.311 & -0.335$\pm$0.029 \\
$[16, 18]$ & -0.229$\pm$0.021 & -3.409$\pm$0.207 & -4.659$\pm$0.283 & -0.433$\pm$0.024 \\
$[18, 20]$ & -0.250$\pm$0.018 & -3.464$\pm$0.173 & -4.720$\pm$0.240 & -0.590$\pm$0.019 \\
$[15, 20]$ & -0.231$\pm$0.019 & -3.420$\pm$0.197 & -4.671$\pm$0.283 & -0.467$\pm$0.024 \\
\bottomrule
\end{tabular}
\end{table}

\subsection{$\A_{f_L}$ (in units of $10^{-2}$) }

\begin{table}[H]
\centering
\begin{tabular}{lllll}
\toprule
Bin & SM & Case I & Case II & Case III \\
\midrule
$[0.1, 1]$ & 0.033$\pm$0.003 & -0.081$\pm$0.043 & -0.130$\pm$0.059 & 0.042$\pm$0.007 \\
$[1, 2]$ & 0.039$\pm$0.011 & -0.244$\pm$0.055 & -0.358$\pm$0.074 & 0.058$\pm$0.018 \\
$[2, 4]$ & 0.068$\pm$0.012 & -0.247$\pm$0.052 & -0.367$\pm$0.073 & 0.087$\pm$0.018 \\
$[4, 6]$ & 0.084$\pm$0.011 & -0.189$\pm$0.053 & -0.284$\pm$0.078 & 0.097$\pm$0.014 \\
$[1.1, 6]$ & 0.071$\pm$0.011 & -0.220$\pm$0.052 & -0.328$\pm$0.075 & 0.087$\pm$0.016 \\
$[15, 16]$ & -0.084$\pm$0.009 & -1.516$\pm$0.098 & -2.076$\pm$0.136 & -0.123$\pm$0.015 \\
$[16, 18]$ & -0.090$\pm$0.008 & -1.419$\pm$0.085 & -1.938$\pm$0.118 & -0.155$\pm$0.013 \\
$[18, 20]$ & -0.090$\pm$0.007 & -1.285$\pm$0.061 & -1.750$\pm$0.083 & -0.202$\pm$0.008 \\
$[15, 20]$ & -0.089$\pm$0.008 & -1.392$\pm$0.079 & -1.901$\pm$0.105 & -0.165$\pm$0.012 \\
\bottomrule
\end{tabular}
\end{table}

\subsection{$\A_{1c}$ (in units of $10^{-2}$) }

\begin{table}[H]
\centering
\begin{tabular}{lllll}
\toprule
Bin & SM & Case I & Case II & Case III \\
\midrule
$[0.1, 1]$ & 0.020$\pm$0.004 & 0.021$\pm$0.004 & -0.055$\pm$0.019 & 0.022$\pm$0.005 \\
$[1, 2]$ & 0.044$\pm$0.010 & 0.048$\pm$0.011 & -0.118$\pm$0.045 & 0.046$\pm$0.010 \\
$[2, 4]$ & 0.044$\pm$0.011 & 0.051$\pm$0.011 & -0.122$\pm$0.064 & 0.046$\pm$0.013 \\
$[4, 6]$ & 0.030$\pm$0.009 & 0.036$\pm$0.010 & -0.100$\pm$0.072 & 0.032$\pm$0.010 \\
$[1.1, 6]$ & 0.037$\pm$0.009 & 0.044$\pm$0.010 & -0.111$\pm$0.065 & 0.040$\pm$0.010 \\
$[15, 16]$ & 0.080$\pm$0.008 & 0.100$\pm$0.010 & 1.213$\pm$0.112 & 0.080$\pm$0.008 \\
$[16, 18]$ & 0.083$\pm$0.007 & 0.103$\pm$0.009 & 1.206$\pm$0.094 & 0.082$\pm$0.007 \\
$[18, 20]$ & 0.069$\pm$0.006 & 0.086$\pm$0.007 & 1.006$\pm$0.074 & 0.069$\pm$0.006 \\
$[15, 20]$ & 0.078$\pm$0.007 & 0.096$\pm$0.008 & 1.137$\pm$0.084 & 0.077$\pm$0.007 \\
\bottomrule
\end{tabular}
\end{table}

\subsection{$\A_{2c}$ (in units of $10^{-2}$) }

\begin{table}[H]
\centering
\begin{tabular}{lllll}
\toprule
Bin & SM & Case I & Case II & Case III \\
\midrule
$[0.1, 1]$ & -0.004$\pm$0.017 & -0.004$\pm$0.017 & 0.048$\pm$0.021 & -0.006$\pm$0.017 \\
$[1, 2]$ & -0.015$\pm$0.029 & -0.015$\pm$0.032 & 0.104$\pm$0.047 & -0.018$\pm$0.029 \\
$[2, 4]$ & -0.025$\pm$0.015 & -0.023$\pm$0.025 & 0.099$\pm$0.053 & -0.026$\pm$0.015 \\
$[4, 6]$ & -0.031$\pm$0.015 & -0.026$\pm$0.010 & 0.068$\pm$0.057 & -0.031$\pm$0.015 \\
$[1.1, 6]$ & -0.026$\pm$0.010 & -0.023$\pm$0.018 & 0.086$\pm$0.052 & -0.027$\pm$0.010 \\
$[15, 16]$ & -0.118$\pm$0.087 & -0.126$\pm$0.075 & -1.03$\pm$0.123 & -0.153$\pm$0.086 \\
$[16, 18]$ & -0.128$\pm$0.094 & -0.138$\pm$0.082 & -1.101$\pm$0.118 & -0.180$\pm$0.094 \\
$[18, 20]$ & -0.138$\pm$0.102 & -0.150$\pm$0.091 & -1.197$\pm$0.120 & -0.224$\pm$0.102 \\
$[15, 20]$ & -0.129$\pm$0.095 & -0.140$\pm$0.086 & -1.12$\pm$0.122 & -0.190$\pm$0.095 \\
\bottomrule
\end{tabular}
\end{table}

\subsection{$\A_{2cc}$ (in units of $10^{-2}$) }

\begin{table}[H]
\centering
\begin{tabular}{lllll}
\toprule
Bin & SM & Case I & Case II & Case III \\
\midrule
$[0.1, 1]$ & -0.076$\pm$0.101 & -0.115$\pm$0.110 & -0.129$\pm$0.103 & -0.074$\pm$0.100 \\
$[1, 2]$ & -0.034$\pm$0.044 & -0.123$\pm$0.061 & -0.158$\pm$0.072 & -0.031$\pm$0.043 \\
$[2, 4]$ & 0.003$\pm$0.024 & -0.089$\pm$0.055 & -0.124$\pm$0.066 & 0.005$\pm$0.024 \\
$[4, 6]$ & 0.025$\pm$0.030 & -0.047$\pm$0.051 & -0.075$\pm$0.067 & 0.026$\pm$0.030 \\
$[1.1, 6]$ & 0.008$\pm$0.029 & -0.075$\pm$0.047 & -0.107$\pm$0.064 & 0.010$\pm$0.029 \\
$[15, 16]$ & 0.092$\pm$0.082 & 0.703$\pm$0.100 & 0.944$\pm$0.114 & 0.092$\pm$0.082 \\
$[16, 18]$ & 0.096$\pm$0.083 & 0.702$\pm$0.099 & 0.941$\pm$0.111 & 0.096$\pm$0.083 \\
$[18, 20]$ & 0.083$\pm$0.069 & 0.589$\pm$0.079 & 0.788$\pm$0.082 & 0.082$\pm$0.069 \\
$[15, 20]$ & 0.091$\pm$0.078 & 0.663$\pm$0.087 & 0.888$\pm$0.104 & 0.090$\pm$0.078 \\
\bottomrule
\end{tabular}
\end{table}

\subsection{$\A_{2ss}$ (in units of $10^{-2}$) }

\begin{table}[H]
\centering
\begin{tabular}{lllll}
\toprule
Bin & SM & Case I & Case II & Case III \\
\midrule
$[0.1, 1]$ & -0.014$\pm$0.111 & 0.001$\pm$0.116 & 0.011$\pm$0.112 & -0.013$\pm$0.110 \\
$[1, 2]$ & 0.040$\pm$0.137 & 0.087$\pm$0.140 & 0.111$\pm$0.139 & 0.043$\pm$0.137 \\
$[2, 4]$ & 0.060$\pm$0.147 & 0.120$\pm$0.140 & 0.147$\pm$0.150 & 0.062$\pm$0.146 \\
$[4, 6]$ & 0.062$\pm$0.142 & 0.120$\pm$0.154 & 0.143$\pm$0.143 & 0.064$\pm$0.142 \\
$[1.1, 6]$ & 0.058$\pm$0.149 & 0.115$\pm$0.148 & 0.139$\pm$0.152 & 0.060$\pm$0.149 \\
$[15, 16]$ & 0.122$\pm$0.111 & 0.958$\pm$0.128 & 1.286$\pm$0.139 & 0.123$\pm$0.112 \\
$[16, 18]$ & 0.119$\pm$0.106 & 0.896$\pm$0.114 & 1.201$\pm$0.120 & 0.121$\pm$0.107 \\
$[18, 20]$ & 0.096$\pm$0.088 & 0.699$\pm$0.090 & 0.934$\pm$0.098 & 0.097$\pm$0.089 \\
$[15, 20]$ & 0.112$\pm$0.097 & 0.840$\pm$0.108 & 1.125$\pm$0.123 & 0.113$\pm$0.097 \\
\bottomrule
\end{tabular}
\end{table}

\subsection{$\A_{3s}$ (in units of $10^{-2}$) }

\begin{table}[H]
\centering
\begin{tabular}{lllll}
\toprule
Bin & SM & Case I & Case II & Case III \\
\midrule
$[0.1, 1]$ & 0.021$\pm$0.023 & 0.022$\pm$0.024 & -0.636$\pm$0.780 & 0.392$\pm$0.061 \\
$[1, 2]$ & 0.026$\pm$0.026 & 0.029$\pm$0.030 & -0.859$\pm$0.959 & -0.194$\pm$0.226 \\
$[2, 4]$ & 0.020$\pm$0.020 & 0.023$\pm$0.023 & -0.680$\pm$0.738 & -1.059$\pm$0.270 \\
$[4, 6]$ & 0.013$\pm$0.012 & 0.015$\pm$0.015 & -0.451$\pm$0.494 & -1.686$\pm$0.196 \\
$[1.1, 6]$ & 0.017$\pm$0.018 & 0.020$\pm$0.021 & -0.604$\pm$0.683 & -1.223$\pm$0.244 \\
$[15, 16]$ & 0.002$\pm$0.002 & 0.002$\pm$0.002 & -0.039$\pm$0.068 & -2.047$\pm$0.038 \\
$[16, 18]$ & 0.001$\pm$0.001 & 0.001$\pm$0.002 & -0.020$\pm$0.056 & -1.917$\pm$0.045 \\
$[18, 20]$ & 0.000$\pm$0.001 & 0.001$\pm$0.001 & -0.002$\pm$0.040 & -1.494$\pm$0.058 \\
$[15, 20]$ & 0.001$\pm$0.001 & 0.001$\pm$0.001 & -0.018$\pm$0.051 & -1.796$\pm$0.046 \\
\bottomrule
\end{tabular}
\end{table}

\subsection{$\A_{3sc}$ (in units of $10^{-2}$) }

\begin{table}[H]
\centering
\begin{tabular}{lllll}
\toprule
Bin & SM & Case I & Case II & Case III \\
\midrule
$[0.1, 1]$ & -0.023$\pm$0.032 & 0.702$\pm$0.853 & 1.003$\pm$1.232 & -0.618$\pm$0.038 \\
$[1, 2]$ & -0.030$\pm$0.038 & 0.918$\pm$1.034 & 1.321$\pm$1.522 & -0.802$\pm$0.069 \\
$[2, 4]$ & -0.022$\pm$0.025 & 0.658$\pm$0.667 & 0.945$\pm$0.992 & -0.600$\pm$0.081 \\
$[4, 6]$ & -0.013$\pm$0.013 & 0.384$\pm$0.333 & 0.547$\pm$0.479 & -0.384$\pm$0.063 \\
$[1.1, 6]$ & -0.019$\pm$0.022 & 0.575$\pm$0.571 & 0.822$\pm$0.826 & -0.530$\pm$0.073 \\
$[15, 16]$ & -0.001$\pm$0.002 & 0.023$\pm$0.033 & 0.032$\pm$0.047 & -0.046$\pm$0.006 \\
$[16, 18]$ & -0.001$\pm$0.001 & 0.014$\pm$0.023 & 0.019$\pm$0.035 & -0.031$\pm$0.004 \\
$[18, 20]$ & -0.000$\pm$0.001 & 0.005$\pm$0.012 & 0.007$\pm$0.017 & -0.013$\pm$0.002 \\
$[15, 20]$ & -0.000$\pm$0.001 & 0.013$\pm$0.021 & 0.017$\pm$0.029 & -0.028$\pm$0.004 \\
\bottomrule
\end{tabular}
\end{table}

\subsection{$\A_{4s}$ (in units of $10^{-2}$) }

\begin{table}[H]
\centering
\begin{tabular}{lllll}
\toprule
Bin & SM & Case I & Case II & Case III \\
\midrule
$[0.1, 1]$ & -0.004$\pm$0.016 & -0.004$\pm$0.017 & 0.031$\pm$0.048 & -0.029$\pm$0.017 \\
$[1, 2]$ & -0.010$\pm$0.016 & -0.008$\pm$0.016 & 0.017$\pm$0.042 & -0.037$\pm$0.019 \\
$[2, 4]$ & -0.011$\pm$0.024 & -0.008$\pm$0.017 & -0.009$\pm$0.038 & -0.027$\pm$0.026 \\
$[4, 6]$ & -0.008$\pm$0.027 & -0.005$\pm$0.022 & -0.020$\pm$0.035 & -0.015$\pm$0.031 \\
$[1.1, 6]$ & -0.010$\pm$0.023 & -0.007$\pm$0.017 & -0.010$\pm$0.040 & -0.023$\pm$0.025 \\
$[15, 16]$ & 0.039$\pm$0.032 & 0.042$\pm$0.029 & 0.356$\pm$0.081 & 0.165$\pm$0.031 \\
$[16, 18]$ & 0.060$\pm$0.049 & 0.065$\pm$0.042 & 0.534$\pm$0.084 & 0.185$\pm$0.047 \\
$[18, 20]$ & 0.100$\pm$0.071 & 0.108$\pm$0.069 & 0.875$\pm$0.088 & 0.225$\pm$0.070 \\
$[15, 20]$ & 0.070$\pm$0.055 & 0.076$\pm$0.044 & 0.617$\pm$0.082 & 0.195$\pm$0.054 \\
\bottomrule
\end{tabular}
\end{table}

\subsection{$\A_{4sc}$ (in units of $10^{-2}$) }

\begin{table}[H]
\centering
\begin{tabular}{lllll}
\toprule
Bin & SM & Case I & Case II & Case III \\
\midrule
$[0.1, 1]$ & 0.003$\pm$0.013 & -0.016$\pm$0.023 & -0.024$\pm$0.032 & 0.028$\pm$0.015 \\
$[1, 2]$ & 0.006$\pm$0.017 & -0.019$\pm$0.033 & -0.029$\pm$0.047 & 0.040$\pm$0.020 \\
$[2, 4]$ & 0.006$\pm$0.019 & -0.011$\pm$0.030 & -0.017$\pm$0.038 & 0.033$\pm$0.020 \\
$[4, 6]$ & 0.005$\pm$0.017 & -0.002$\pm$0.027 & -0.005$\pm$0.030 & 0.023$\pm$0.019 \\
$[1.1, 6]$ & 0.005$\pm$0.017 & -0.008$\pm$0.027 & -0.013$\pm$0.033 & 0.029$\pm$0.018 \\
$[15, 16]$ & 0.010$\pm$0.011 & 0.087$\pm$0.034 & 0.117$\pm$0.046 & 0.015$\pm$0.011 \\
$[16, 18]$ & 0.011$\pm$0.011 & 0.088$\pm$0.028 & 0.118$\pm$0.037 & 0.014$\pm$0.011 \\
$[18, 20]$ & 0.009$\pm$0.009 & 0.075$\pm$0.020 & 0.100$\pm$0.025 & 0.011$\pm$0.009 \\
$[15, 20]$ & 0.010$\pm$0.011 & 0.083$\pm$0.026 & 0.112$\pm$0.034 & 0.013$\pm$0.011 \\
\bottomrule
\end{tabular}
\end{table}

\bibliography{bibliography}{}
\bibliographystyle{utcaps_mod}
\end{document}